# Housing Price Appreciation, Housing Affordability, and the Spatial Restructuring of Toronto Commuting

**Shawn Berry, DBA[1*]**

**September 4, 2026**

[1]William Howard Taft University, Lakewood CO, USA

*Correspondence: shawnpberry@gmail.com

---

**Abstract**

This study investigates the spatial relationships among housing-market appreciation, housing affordability, and automobile commuting to Toronto across 23 census metropolitan areas and municipalities in southern Ontario from 2016 to 2021. Utilizing Statistics Canada Census data, the analysis employed linear and quadratic ordinary least-squares regression, Global and Local Moran's I, and residual spatial diagnostics to assess geographic patterns in average dwelling values, owner shelter-cost burden, and lone-driver commuting. The study found that average dwelling values increased significantly throughout the region, maintaining a strong nonlinear relationship with distance from Toronto, which indicates a persistent metropolitan accessibility gradient. Commuter volumes also demonstrated pronounced nonlinear distance gradients; however, housing appreciation was not significantly correlated with percentage changes in commuting ($R^2 = .045$, $p = .331$), suggesting that the expansion of the housing market did not result in proportional growth in Toronto-bound automobile travel. The shelter-cost burden exhibited weaker distance relationships and greater residual spatial dependence, indicating localized affordability influences not fully explained by distance or dwelling values. Overall, the findings suggest an increasingly extensive regional housing market with spatially heterogeneous affordability and commuting outcomes, highlighting the importance of integrating housing, transportation, and regional planning.

## 1. Introduction

### 1.1 *Background and Rationale*

Housing affordability has become a defining economic and social challenge in large metropolitan regions. Over the past decade, many cities have experienced unprecedented increases in residential property values, reducing access to home ownership, and altering the spatial distribution of households. In high-growth metropolitan areas, rapidly appreciating housing markets increasingly compel households to balance competing objectives: remaining close to employment centers while seeking housing that is financially attainable. This trade-off has important implications not only for residential location decisions but also for commuting behavior, transportation demand, and the long-term spatial organization of metropolitan regions. The Greater Toronto Area (GTA) provides an important setting for examining these dynamics. Between 2016 and 2021, residential property values increased dramatically across southern Ontario, with many communities experiencing appreciation exceeding 70% and some nearly doubling in value. The average shelter costs also increased, although at a considerably slower pace. During the same period, commuting patterns evolved in response to multiple structural forces, including rapid housing market appreciation, changing labor market conditions, regional economic growth, and the widespread adoption of remote and hybrid work arrangements associated with the COVID-19 pandemic. Together, these developments have created an opportunity to investigate how rapidly changing housing markets influence the geography of automobile commuting.

Urban economic theory suggests that households continuously evaluate the trade-off between housing and commuting costs when selecting residential locations. Classical bid-rent and monocentric city models propose that households can reduce housing expenditures by locating farther from employment centers while accepting longer commuting distances. As housing prices increase within metropolitan cores, the economic incentive to relocate to lower-cost peripheral communities becomes stronger, potentially expanding the metropolitan commuter shed. However, this relationship is unlikely to be purely linear. Housing markets exhibit complex spatial gradients, transportation accessibility varies across communities, and employment opportunities increasingly extend beyond traditional central business districts. Consequently, understanding how housing appreciation influences commuting requires an empirical examination of nonlinear spatial relationships rather than assuming constant rates of change.
Although a substantial body of literature has examined housing affordability and residential mobility, comparatively fewer studies have investigated how the rapid appreciation of residential property values reshapes commuting geography across an entire metropolitan region. Much of the existing literature emphasizes either commuting behavior or housing markets independently, despite the fact that both decisions are jointly determined by household location choice. Furthermore, recent housing market conditions following the COVID-19 pandemic present a unique opportunity to examine whether accelerating housing prices altered the spatial extent of metropolitan labor markets.

This study investigates the relationship between housing market appreciation, housing affordability, and automobile commuting across communities surrounding Toronto between 2016 and 2021. Using Statistics Canada census data for census metropolitan areas and municipalities across southern Ontario, the analysis evaluates three interrelated spatial relationships: (1) the association between distance from Toronto and average dwelling values, (2) the relationship between distance and owner shelter cost burden, and (3) changes in the number of lone-driver commuters travelling to Toronto. Both linear and quadratic regression models were estimated to determine whether these relationships were best characterized by linear or nonlinear spatial gradients. Particular attention is paid to whether rapid increases in residential property values correspond to evidence of outward residential decentralization and changes in commuter distribution.

This study makes three main contributions. First, it provides empirical evidence that the relationship between metropolitan distance and housing values is strongly nonlinear and that the housing value gradient intensified substantially between 2016 and 2021. Second, it demonstrates that changes in average dwelling values appear to have been more strongly associated with spatial restructuring than changes in shelter costs, suggesting that housing purchase prices exerted a greater influence on residential location decisions than monthly housing expenditures. Third, by jointly examining housing values, affordability, and commuter behavior within a common spatial framework, the study illustrates how rapid housing market appreciation may contribute to the expansion of metropolitan commuter regions while simultaneously reinforcing regional differences in housing affordability.

The remainder of this paper is organized as follows. This paper begins by describing the study area, data sources, and statistical methods used in the analysis. It then presents the empirical findings for automobile commuting, average dwelling values, and owner shelter cost burden before discussing their implications for residential decentralization, housing affordability, and metropolitan commuting patterns. The paper concludes by considering broader policy implications, study limitations, and opportunities for future research. The final section considers the policy implications, study limitations, and directions for future research.

### 1.2 Literature Review

#### 1.2.1. *Urban Spatial Equilibrium and the Housing–Commuting Trade-Off*

Urban spatial-equilibrium theory conceptualizes residential location as a joint decision involving housing prices, wages, accessibility, commuting time, and transportation costs, rather than as an isolated housing choice (So et al., 2001). Households may obtain access to higher metropolitan wages and employment opportunities while residing in less expensive peripheral housing markets, but this strategy requires them to absorb commuting costs that generally rise with distance from the employment center (So et al., 2001). Empirical research on joint residential and workplace choices confirms that households trade off housing prices, wages, and travel time when deciding where to live and work (So et al., 2001). The theoretical relevance of this trade-off to the Toronto region is that communities outside the metropolitan core may remain economically integrated with Toronto, even when they constitute separate municipal housing markets (So et al., 2001; Roberts & Taylor, 2017).

Commuting connects geographically separated housing and labor markets by allowing households to reside outside the municipality in which they work (Roberts & Taylor, 2017; So et al., 2001). Labor-market research commonly assumes that longer commutes must be compensated through higher wages, lower housing costs, improved amenities, or other favorable employment and residential conditions (Roberts & Taylor, 2017). This compensating-differential framework implies that housing affordability cannot be assessed independently of the monetary and time costs required to obtain employment opportunities (Roberts & Taylor, 2017; So et al., 2001). Therefore, a household may accept a longer commute when the housing savings or residential amenities obtained at a peripheral location exceed the additional generalized cost of travel (So et al., 2001).

The housing–commuting trade-off does not require dwelling values or commuting volumes to change at a constant rate with distance (Dubin and Sung 1987). Urban housing markets are influenced by multiple employment centers, transportation corridors, neighborhood amenities, municipal boundaries, and local differences in housing supply, all of which can generate spatially varying price gradients (Dubin & Sung, 1987). Empirical analysis of non-monocentric cities has demonstrated that housing price gradients can differ across space because accessibility is not determined solely by distance from one central business district (Dubin & Sung, 1987). These findings support the use of flexible nonlinear specifications when studying a large and increasingly polycentric metropolitan region, such as Toronto (Dubin & Sung, 1987).

#### 1.2.2. *Housing-Value Gradients and Metropolitan Accessibility*

Housing-value gradients provide an empirical representation of the value placed on metropolitan accessibility because employment access, transportation infrastructure, and urban amenities may be capitalized into property prices (Dubin & Sung, 1987; So et al., 2001). When access to employment and services is concentrated near a metropolitan center, housing values are expected to be higher in accessible locations and lower in places that require longer journeys to the center (Dubin & Sung, 1987; So et al., 2001). The magnitude of the gradient may vary across metropolitan space because accessibility, housing quality, amenities, and employment density are not distributed uniformly (Dubin & Sung, 1987). Consequently, a single linear coefficient may conceal curvature or other spatial heterogeneity in the relationship between distance and dwelling values (Dubin & Sung, 1987).

A negative first-order distance coefficient combined with a positive quadratic coefficient describes a curve in which dwelling values initially decline rapidly with distance, but the marginal decline becomes smaller farther from the metropolitan center (Dubin & Sung, 1987). Such a quadratic pattern is consistent with a strong proximity premium near the core and a weaker marginal effect of additional distance among peripheral communities (Dubin & Sung, 1987). The presence of a positive quadratic term does not by itself establish that dwelling values genuinely rise in extremely distant markets, particularly when the estimated turning point lies beyond the observed data range (Akaike, 1974). Extrapolation beyond the sampled range should therefore be treated as a mathematical implication of the fitted curve rather than direct evidence of the behavior of unobserved housing markets (Akaike, 1974; Ramsey, 1969).

Comparing housing value gradients across time can reveal whether metropolitan housing demand has spread outward while preserving the relative value of accessibility (Dubin & Sung, 1987; Van Nieuwerburgh, 2023). An upward shift in the entire distance–value curve indicates broad housing appreciation, whereas an outward shift in the curve’s stationary point suggests that the metropolitan price influence extends farther into peripheral communities (Dubin & Sung, 1987). Remote-work research further indicates that reduced commuting frequency can increase residential demand for peripheral locations and alter relative housing values across urban spaces (Delventhal et al., 2022; Davis et al., 2024; Van Nieuwerburgh, 2023). The comparison between 2016 and 2021 can therefore be interpreted as reflecting both conventional accessibility capitalization and changing workplace-location constraints (Delventhal et al., 2022; Davis et al., 2024; Van Nieuwerburgh, 2023).

### 1.2.3. *Housing Appreciation, Affordability, and Residential Sorting*

Housing appreciation can influence residential sorting by changing which households purchase homes in high-value metropolitan submarkets (So et al. 2001). When dwelling values rise more rapidly near major employment centers, prospective buyers may seek housing farther away while attempting to retain access to metropolitan employment (So et al., 2001). Joint residential and job location evidence indicates that households respond to differences in wages, housing prices, and commuting time rather than to any one of these factors independently (So et al., 2001). Residential decentralization may therefore represent an affordability response, even when households continue to maintain economic ties with the central metropolitan labor market (So et al., 2001; Roberts & Taylor, 2017).

The average dwelling value and average shelter cost measure different dimensions of housing affordability (So et al., 2001; Van Nieuwerburgh, 2023). The dwelling value represents the market price of the residential asset and, therefore, more closely reflects the financial entry barrier encountered by prospective purchasers (Van Nieuwerburgh, 2023). Current owner shelter costs depend on mortgage balances, interest rates, property taxes, utilities, and the timing of acquisition, which means that households occupying similarly valued properties may face substantially different monthly expenditures (Van Nieuwerburgh, 2023). Therefore, existing owners can experience substantial capital appreciation without a proportional increase in current monthly shelter payments, whereas new purchasers must finance housing at prevailing market prices (Van Nieuwerburgh, 2023).

This distinction provides a theoretical basis for dwelling values to exhibit a stronger relationship with residential location than with contemporaneous shelter costs (So et al., 2001; Van Nieuwerburgh, 2023). Purchase prices directly affect mortgage qualification, required down payments, and the feasibility of entering particular housing markets, while shelter costs among existing owners partly reflect historical financing arrangements (Van Nieuwerburgh, 2023). Therefore, a large increase in dwelling values accompanied by a smaller change in average shelter costs is compatible with an increasingly restrictive ownership market rather than an improvement in affordability for new entrants (Van Nieuwerburgh, 2023). In this context, outward residential movement may be driven more strongly by the capital cost of purchasing housing than by the average monthly expenditure of households that already own them (So et al., 2001; Van Nieuwerburgh, 2023).

### 1.2.4. *Commuting as an Indicator of Functional Metropolitan Integration*

Commuter flows provide evidence of functional economic integration because they connect places of residence to geographically separate employment markets (Roberts & Taylor, 2017; So et al., 2001). A municipality can be located outside Toronto while remaining part of Toronto's functional economic region if a meaningful share of residents travel to Toronto for employment (Roberts & Taylor, 2017). Distance is expected to constrain these flows because commuting requires both travel time and direct transportation expenditure (Roberts & Taylor, 2017; So et al., 2001). Nevertheless, the effect of distance may be nonlinear because commuter volumes are likely to be concentrated in nearby metropolitan areas and substantially smaller locations lie beyond the principal commuter belt (Dubin & Sung, 1987; Roberts & Taylor, 2017). Observed commuter counts also depend on municipal population, labor-force size, local employment, transportation access, and the historical development of settlement patterns (Roberts & Taylor, 2017). Consequently, large municipalities can generate more commuters than small municipalities, even when their residents have comparable propensities to travel to Toronto (Roberts & Taylor, 2017). Raw commuter counts, therefore, measure the absolute scale of the commuting relationship rather than an individual probability of commuting (Cameron & Trivedi, 1986). Rate-based measures, such as Toronto-bound commuters per employed resident, would answer a different question by adjusting for the population at risk of commuting (Cameron & Trivedi, 1986).

The extremely large commuter count for Toronto relative to other observations is substantively meaningful but statistically influential (Cook, 1977). Toronto represents the metropolitan core and has a labor force and population that are not comparable in scale to those of smaller surrounding municipalities (Cook, 1977). Including Toronto allows the regression to describe the full regional distribution of commuters, but it may cause the fitted coefficients and confidence intervals to depend heavily on one observation (Cook, 1977). Sensitivity analyses that exclude Toronto or use robust and transformed models are therefore necessary to determine whether the estimated relationship characterizes the wider regional pattern or primarily reflects the core–periphery contrast (Cook, 1977; Huber, 1964).

### 1.2.5. *Remote Work and the Reorganization of Metropolitan Space*

The COVID-19 pandemic has produced a substantial expansion of remote and hybrid work, thereby weakening the requirement that workers reside near their workplaces (Van Nieuwerburgh, 2023). Research on work from home identifies the consequences of residential location, traffic congestion, commercial property, housing demand, and the spatial organization of cities (Delventhal et al., 2022; Davis et al., 2024; Van Nieuwerburgh, 2023). A quantitative urban model developed by Delventhal et al. (2022) predicts that increased work from home can move residents toward the urban periphery while changing congestion and travel times. Equilibrium modelling by Davis et al. (2024) similarly treats residential location, workplace attendance, and demand for space as jointly determined responses to work-from-home technology.

Remote work can produce residential decentralization without a corresponding increase in the number of daily commuting trips (Davis et al., 2022; Davis et al., 2024). A household may relocate farther from Toronto because fewer required office days reduce the effective weekly cost of a long commute (Davis et al., 2024; Van Nieuwerburgh, 2023). The resulting household can remain economically connected to Toronto while appearing less frequently in journey-to-work measures that capture usual commuting behavior (Davis et al., 2024). This mechanism explains why rapid peripheral housing appreciation and outward residential movement may coexist with declining or heterogeneous commuter counts between 2016 and 2021 (Delventhal et al., 2022; Van Nieuwerburgh, 2023). Remote work is not distributed uniformly across occupations, workers, or locations; therefore, its spatial effects are likely to differ across communities (Davis et al. 2024; Van Nieuwerburgh 2023). Communities with substantial populations of telework-compatible workers may experience increased housing demand without a proportionate increase in daily automobile commutes (Davis et al. 2024). Larger established commuter markets may also record reductions in observed trips if workers shift from daily to hybrid attendance while retaining the same residence and employer (Van Nieuwerburgh, 2023). Commuting changes during the study period should consequently be interpreted as the joint outcome of housing prices, residential relocation, workplace flexibility, and local labor-market conditions rather than as a direct response to dwelling values alone (Delventhal et al., 2022; Davis et al., 2024; Van Nieuwerburgh, 2023).

### 1.3 Research objectives

The objective of this study is to examine the relationships among distance from Toronto, average dwelling values, owner housing affordability, and Toronto-bound automobile commuting across municipalities and census metropolitan areas in southern Ontario using the 2016 and 2021 Census of Canada data. Specifically, this study (1) quantifies the spatial relationships between distance from Toronto and average dwelling values, owner shelter-cost burden, and commuter volumes; (2) evaluates the extent to which average dwelling values are associated with the proportion of owner households spending more than 30% of household income on shelters; (3) compares linear and quadratic regression models to determine the functional form that best characterizes these relationships; and (4) assesses how these relationships changed between 2016 and 2021 in the context of widespread housing price appreciation and evolving commuting behavior. By integrating measures of housing values, housing affordability, and commuting within a common spatial framework, this study contributes new empirical evidence regarding the regional restructuring of housing markets and travel behavior in Canada's largest metropolitan region.

### 1.4 Significance and Contribution

The literature supports the central theoretical premise that housing prices, residential location, and commuting are jointly determined within the metropolitan spatial equilibrium (Roberts & Taylor, 2017; So et al., 2001). Research on non-monocentric housing markets supports nonlinear distance gradients because accessibility and employment are distributed unevenly across metropolitan space (Dubin & Sung, 1987). Research on remote work indicates that workplace flexibility can extend residential location choices outward while simultaneously reducing commuting frequency (Delventhal et al., 2022; Davis et al., 2024; Van Nieuwerburgh, 2023). These bodies of evidence jointly support the examination of changes in dwelling values and commuter volumes within a common Toronto-centered spatial framework (Delventhal et al., 2022; So et al., 2001).

The empirical strategy is supported by a descriptive, comparative approach, rather than a causal design (Akaike, 1974; Ramsey, 1969). Linear and quadratic comparisons test whether the distance gradient is constant or curved, while adjusted $R^2$, information criteria, RMSE, coefficient tests, and confidence intervals provide complementary evidence regarding model performance (Akaike, 1974). Shapiro–Wilk, and Moran's *I* address distinct assumptions concerning distribution, variance, specification, influence, and spatial dependence (Breusch & Pagan, 1979; Cook, 1977; Moran, 1950; Ramsey, 1969; Shapiro & Wilk, 1965).

The principal contribution of this study is the integration of housing-value gradients and automobile commuting within the same regional and temporal analysis (So et al., 2001). The findings can identify whether Toronto's housing-price influence extended outward and whether the absolute distribution of Toronto-oriented commuters changed during the same period (So et al. 2001; Van Nieuwerburgh 2023). This study cannot identify the individual motives underlying relocation, but it can document the spatial structure within which housing affordability, residential decentralization, and commuting decisions occur (Delventhal et al., 2022; So et al., 2001). This contribution is particularly relevant for integrated housing and transportation policies because housing savings achieved through peripheral relocation may be accompanied by greater transportation dependence or altered workplace access requirements (Roberts & Taylor, 2017; So et al., 2001).

## 2. Materials and Methods

This study employed a comparative, cross-sectional ecological design using publicly available data from the 2016 and 2021 Census of Canada for selected census metropolitan areas and municipalities at varying distances from Toronto. Commuting was measured as the number of employed residents who reported driving to a workplace in Toronto, while housing-market conditions were represented by average dwelling value and the percentage of owner households spending 30% or more of household income on shelter costs. Distance from Toronto served as the principal spatial predictor, and variables were standardized across census years to facilitate temporal comparison. Descriptive statistics and Shapiro–Wilk tests assessed distributional characteristics and normality. Relationships among distance, commuter volumes, dwelling values, and shelter-cost burden were examined using scatterplots and linear and quadratic ordinary least-squares (OLS) regression. Because linear regression assumes a constant marginal association, quadratic specifications were used where theory and visual evidence suggested that spatial gradients varied with distance (Dubin & Sung, 1987; Ramsey, 1969). Distance was centered before constructing the quadratic term to reduce nonessential collinearity between polynomial components. Spatial dependence was assessed using Global Moran's *I* with 999 permutation tests based on symmetrized, row-standardized four-nearest-neighbor (k = 4) spatial weights; inverse-distance weights were used as a robustness specification, and Moran's *I* was also applied to regression residuals to determine whether spatial autocorrelation remained after accounting for the modeled relationships. Local Indicators of Spatial Association (LISA) were additionally estimated using 999 conditional permutations to identify localized High–High, Low–Low, High–Low, and Low–High spatial clusters (Anselin, 1995). All statistical tests were two-tailed with $\alpha = .05$.

Linear and quadratic models were compared using $R^2$, adjusted $R^2$, RMSE, AIC, BIC, overall *F* tests, coefficient significance, 95% confidence intervals, and graphical fit. Because additional predictors can mechanically increase $R^2$, model selection emphasized measures that account for model complexity and predictive error rather than unadjusted $R^2$ alone (Akaike, 1974). Quadratic models were interpreted as parsimonious representations of spatial curvature rather than evidence that the underlying relationships were inherently polynomial (Akaike, 1974; Ramsey, 1969). Where applicable, stationary points were used to characterize changes in the fitted spatial gradient, but were considered substantively meaningful only when supported by observations within the study range; points outside that range were treated as extrapolations (Akaike, 1974; Ramsey, 1969). Finally, because the observations represent aggregate geographic units, all estimated relationships were interpreted as municipality-level spatial associations rather than causal effects or evidence of individual household relocation behavior.

## 2.1. *Conceptual Framework*

Figure 1 illustrates the conceptual framework. The framework summarizes the study's central premise that increasing dwelling values associated with metropolitan housing appreciation influence residential location decisions, increase the proportion of households experiencing shelter-cost burden, and contribute to the outward redistribution of Toronto-oriented commuters. The observed commuting relationship was moderated by changing workplace practices, including remote and hybrid work, during the 2016–2021 study period.

**Figure 1. Conceptual framework illustrating the relationships between distance from Toronto, average dwelling value, owner shelter cost burden, and Toronto-bound automobile commuting.**

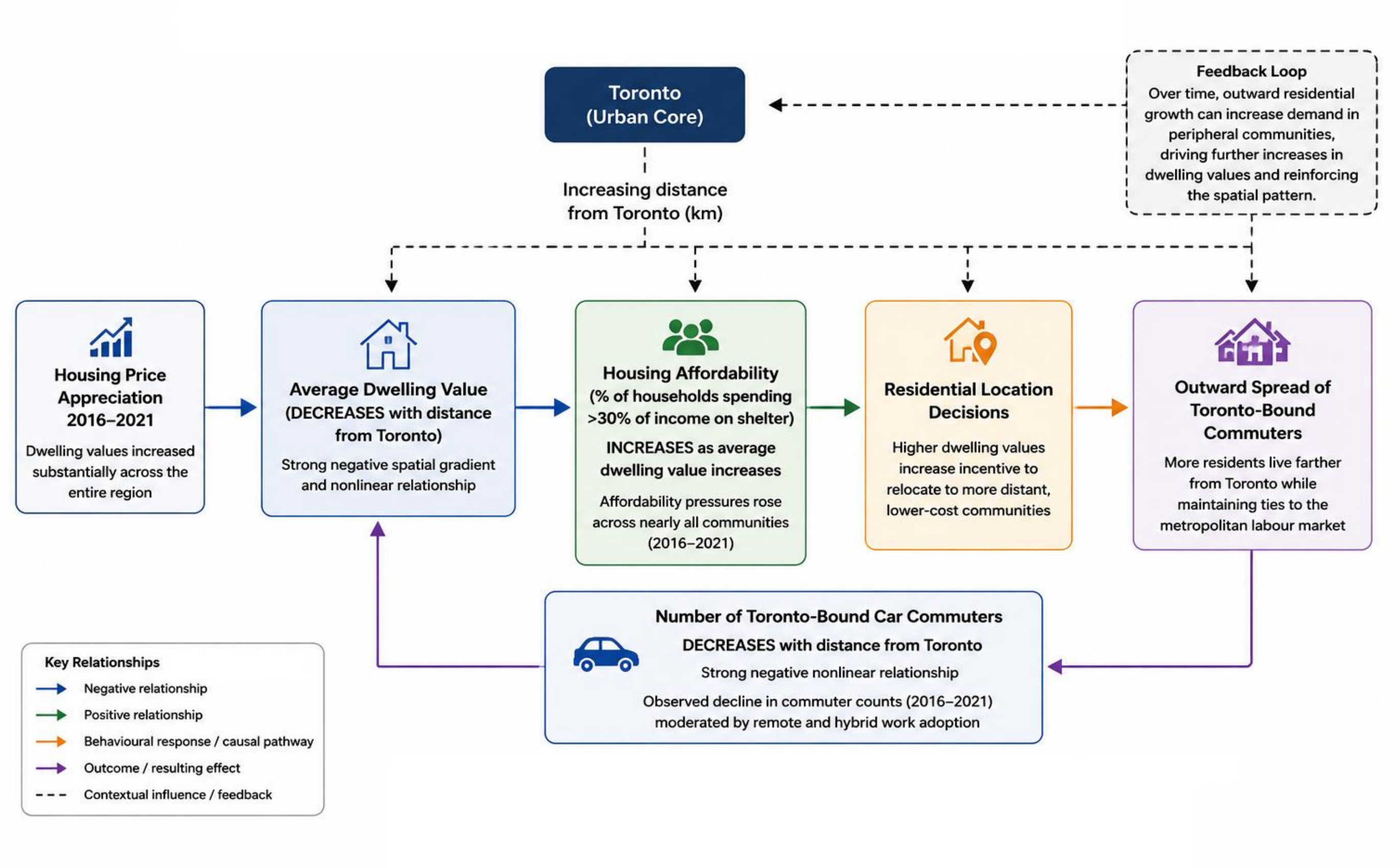

## 3. Results

The sample data were analyzed.

### 3.1. *Dataset*

Table 1 presents the descriptive characteristics of the study municipalities, including their distance from Toronto, average monthly shelter costs, average dwelling values, and the number of Toronto-bound lone-driver commuters in 2016 and 2021. The municipalities span a broad geographic range, from Toronto (20 km) to London (191 km), providing substantial spatial variation for examining housing market conditions and commuting patterns across southern Ontario. Toronto exhibited the highest average dwelling values in both census years, increasing from $734,924 in 2016 to $1,131,000 in 2021, while also recording the highest average shelter costs ($1,755 and $1,948, respectively). In comparison, more distant communities generally reported substantially lower dwelling values and shelter costs, although all municipalities experienced considerable housing appreciation between 2016 and 2021. The magnitude of this appreciation was particularly notable in communities located beyond the Greater Toronto Area, suggesting that rapid increases in residential property values extended well beyond the metropolitan core. These descriptive patterns are consistent with the hypothesis that Toronto's housing market exerts influence across a broad regional housing system rather than being confined to the city itself.

Table 1 also illustrates the considerable heterogeneity in Toronto-oriented commuting. As expected, Toronto recorded the largest number of lone-driver commuters because of its size and employment concentration, with commuter counts declining from 1,274,835 in 2016 to 921,960 in 2021. Among the surrounding municipalities, Oshawa and Hamilton generated the largest commuter flows, whereas more distant communities generally exhibited substantially smaller numbers of Toronto-bound commuters. Nevertheless, the commuter volumes did not decline uniformly with distance. Several municipalities, including Centre Wellington, Orillia, Peterborough, Woodstock, Stratford, Ingersoll, and Belleville–Quinte West, experienced modest increases in Toronto-bound commuters despite being located more than 100 km from Toronto. These observations suggest that commuting behavior is influenced by factors beyond geographic distance alone, including municipal population size, transportation accessibility, regional employment opportunities, and changing workplace arrangements.

The descriptive statistics further revealed important differences in the rates of change among the study variables. While average dwelling values increased dramatically across virtually every municipality, increases in average monthly shelter costs were comparatively modest. At the same time, many of the largest established commuter markets—including Toronto, Oshawa, Hamilton, Barrie, Kitchener–Cambridge–Waterloo, and London—experienced reductions in Toronto-bound commuting between 2016 and 2021. These contrasting trends suggest that the substantial escalation in housing values did not translate directly into proportional increases in commuter volume. Instead, the evidence points toward a more complex process of metropolitan restructuring in which rapid housing appreciation, changing housing affordability, evolving labor markets, and the widespread adoption of remote and hybrid work likely interact to influence residential location and commuting behavior. Consequently, Table 1 provides preliminary evidence supporting the need for subsequent regression analyses to determine whether these apparent spatial patterns represent statistically significant relationships and whether nonlinear distance gradients better characterize housing and commuting dynamics than simple linear trends.

**Table 1. 2016 and 2021 average dwelling values, lone car commuters to Toronto, and percentage of households spending 30% or more of income on shelter.**

| Distance to Toronto (km) | Centre | 2016 Avg value of dwelling | 2021 Avg value of dwelling | 2016 Car commuters – Driver alone | 2021 Car commuters – Driver alone | 2016 Percentage of households spending 30%+ of income on shelter | 2021 Percentage of households spending 30%+ of income on shelter |
|---|---|---|---|---|---|---|---|
| 20 | Toronto CMA | 734,924 | 1,131,000 | 1274835 | 921,960 | 26.7 | 30.5 |
| 60 | Oshawa CMA | 461,757 | 788,000 | 45770 | 35,025 | 18 | 23.5 |
| 68 | Hamilton CMA | 485,415 | 826,000 | 48185 | 31,170 | 16.8 | 23 |
| 93 | Guelph CMA | 463,729 | 789,000 | 6945 | 5,435 | 16.1 | 22 |
| 103 | Brantford CMA | 337,344 | 649,000 | 1860 | 1,690 | 14.8 | 20.5 |
| 106.1 | Port Hope | 352,427 | 669000 | 565 | 515 | 16.8 | 20.9 |
| 106.4 | Centre Wellington | 425,809 | 783,000 | 840 | 895 | 14.1 | 16.9 |
| 107 | Kitchener-Cambridge-Waterloo CMA | 393,896 | 731,000 | 9145 | 7185 | 14.2 | 22 |
| 112 | St. Catharines-Niagara CMA | 309,957 | 616,000 | 4165 | 3375 | 16.8 | 21.7 |
| 113 | Barrie CMA | 419,554 | 746,000 | 18350 | 15295 | 20.3 | 25.3 |
| 117.1 | Cobourg | 342,139 | 616,500 | 330 | 280 | 16.2 | 22.8 |
| 124.9 | Kawartha Lakes | 347,722 | 626,500 | 2790 | 2600 | 17.7 | 19.1 |
| 130.1 | Orillia | 302,658 | 582,000 | 390 | 440 | 18.9 | 26.7 |
| 138 | Peterborough CMA | 337,069 | 619,000 | 1130 | 1205 | 16.2 | 22.1 |
| 143.2 | Woodstock | 266,501 | 563,000 | 230 | 380 | 12.5 | 21.6 |
| 145.7 | Midland | 274,162 | 544,000 | 385 | 220 | 18.7 | 22.4 |
| 147.4 | Collingwood | 390,393 | 778,000 | 400 | 420 | 23.7 | 26.5 |
| 149.9 | Stratford | 316,340 | 546,500 | 25 | 80 | 13.9 | 16.2 |
| 156 | Ingersoll | 246,313 | 511,600 | 10 | 85 | 12.4 | 11 |
| 158.2 | Norfolk | 297,434 | 567,500 | 395 | 245 | 14.7 | 15.6 |
| 189 | Belleville - Quinte West CMA | 252070 | 468400 | 35 | 420 | 14.5 | 18.6 |
| 190.1 | Owen Sound | 289339 | 447200 | 205 | 165 | 14.5 | 20.3 |
| 191 | London CMA | 301631 | 586000 | 2090 | 1075 | 14.2 | 22.3 |

Source: Statistics Canada. Census of Population.

Table 2 illustrates the percentage changes in shelter costs, average dwelling values, and the number of lone car commuters from 2016 to 2021. The data reveal that average dwelling values experienced significant increases across all study areas during this period, while changes in Toronto-bound lone-car commuting exhibited considerable variability. The appreciation in dwelling values ranged from 53.9% in the Toronto CMA to 111.3% in Woodstock, demonstrating that rapid housing-price growth was not limited to Toronto or its immediately adjacent municipalities. Several communities located more than 100 km from Toronto reported increases exceeding 80% or 90%, including Brantford (92.4%), Port Hope (89.8%), St. Catharines–Niagara (98.7%), Orillia (92.3%), Midland (98.4%), Collingwood (99.3%), Ingersoll (107.7%), Norfolk (90.8%), and London (94.3%). This trend aligns with a geographically extensive housing-market expansion, where substantial price appreciation extended well beyond the metropolitan core.

Alterations in lone-car commuting did not exhibit a consistent pattern. Commuting to Toronto decreased in several areas, including Toronto (−27.7%), Oshawa (−23.5%), Hamilton (−35.3%), Guelph (−21.7%), and Kitchener–Cambridge–Waterloo (−21.4%), despite significant increases in property values. At greater distances, the pattern was more variable: Woodstock experienced a 65.2% increase, Stratford 220.0%, Ingersoll 750.0%, and Belleville–Quinte West 1,100.0%, while Midland (−42.9%), Norfolk (−38.0%), and London (−48.6%) saw notable declines. The substantial percentage increases in some peripheral communities should be interpreted with caution, as they reflect changes from relatively small commuter bases in 2016; thus, even modest increases in commuter numbers can result in large percentage changes.

**Table 2. Percentage change in average dwelling value and lone car commuters to Toronto, 2016 to 2021.**

| Distance to Toronto (km) | Centre | % Change Avg Dwelling Value (2016-2021) | % Change Car Commuters (2016-2021) |
|---|---|---|---|
| 20 | Toronto CMA | 53.9% | -27.7% |
| 60 | Oshawa CMA | 70.7% | -23.5% |
| 68 | Hamilton CMA | 70.2% | -35.3% |
| 93 | Guelph CMA | 70.1% | -21.7% |
| 103 | Brantford CMA | 92.4% | -9.1% |
| 106.1 | Port Hope | 89.8% | -8.8% |
| 106.4 | Centre Wellington | 83.9% | 6.5% |
| 107 | Kitchener-Cambridge-Waterloo CMA | 85.6% | -21.4% |
| 112 | St. Catharines-Niagara CMA | 98.7% | -19.0% |
| 113 | Barrie CMA | 77.8% | -16.6% |
| 117.1 | Cobourg | 80.2% | -15.2% |
| 124.9 | Kawartha Lakes | 80.2% | -6.8% |
| 130.1 | Orillia | 92.3% | 12.8% |
| 138 | Peterborough CMA | 83.6% | 6.6% |
| 143.2 | Woodstock | 111.3% | 65.2% |
| 145.7 | Midland | 98.4% | -42.9% |
| 147.4 | Collingwood | 99.3% | 5.0% |
| 149.9 | Stratford | 72.8% | 220.0% |
| 156 | Ingersoll | 107.7% | 750.0% |
| 158.2 | Norfolk | 90.8% | -38.0% |
| 189 | Belleville - Quinte West CMA | 85.8% | 1100.0% |
| 190.1 | Owen Sound | 54.6% | -19.5% |
| 191 | London CMA | 94.3% | -48.6% |

Source: data analysis

Table 3 displays the OLS regression analysis of the percentage change in lone car commuters to Toronto in relation to the percentage change in average dwelling values from 2016 to 2021. The analysis reveals the intricate relationship between housing price appreciation and variations in automobile commuting to Toronto. Despite a general increase in housing values, communities with similar appreciation rates often exhibited markedly different commuting patterns. This interpretation is consistent with the regression analysis of these percentage changes, which demonstrated that the percentage change in dwelling-value appreciation was not a statistically significant predictor of the percentage change in Toronto car commuters, $B = 3.98$, $SE = 4.00$, $t(21) = 0.99$, $p = .331$, 95% CI [−4.34, 12.30], explaining only 4.5% of the observed variation, $R^2 = .045$. Therefore, the findings do not support the conclusion that greater housing appreciation was independently associated with proportionately greater growth or decline in single-car commuting.

**Table 3. OLS regression: Percentage change in lone car commuters to Toronto versus percentage change in average dwelling value, 2016 to 2021.**

| Statistic | OLS Regression |
|---|---|
| Constant, B | -257.601 |
| Constant, SE | 343.064 |
| Constant, t | -0.751 |
| Constant, p | 0.461 |
| Constant, 95% CI lower | -971.042 |
| Constant, 95% CI upper | 455.840 |
| Avg. dwelling value change (%), B | 3.979 |
| Avg. dwelling value change (%), SE | 4.000 |
| Avg. dwelling value change (%), t | 0.995 |
| Avg. dwelling value change (%), p | 0.331 |
| Avg. dwelling value change (%), 95% CI lower | -4.339 |
| Avg. dwelling value change (%), 95% CI upper | 12.297 |
| Model F | 0.990 |
| Numerator df | 1 |
| Denominator df | 21 |
| Model p | 0.331 |
| $R^2$ | 0.045 |
| Adjusted $R^2$ | 0.000 |
| Residual standard error | 277.494 |
| RMSE (in-sample) | 265.155 |
| AIC | 325.966 |
| BIC | 328.237 |
| Residual sum of squares | 1,617,065.748 |
| N | 23 |

Source: data analysis

### 3.2. *Normality tests*

Table 4 presents the descriptive statistics and results of the Shapiro–Wilk test. These results reveal significant differences in the distributional form among the variables employed in the linear and quadratic regression analyses. The variable "Distance from Toronto" was approximately normally distributed, with a mean of 124.74 km, a median of 124.90 km, and a standard deviation of 41.69. The Shapiro–Wilk test did not indicate any deviation from normality ($W = .959$, $p = .435$). This finding suggests that the distribution of the primary geographic predictor was relatively symmetric across the 23 study areas. It is crucial to note, however, that the normality of an independent variable is not a prerequisite for ordinary least squares (OLS) regression. Instead, conventional regression inference primarily focuses on the distribution and behavior of the model residuals, along with assumptions regarding linearity or correct functional form, independence, homoscedasticity, and influential observations (Fox, 2016; Wooldridge, 2020).

In both census years, average dwelling values exhibited significant deviations from normality. In 2016, the average dwelling value was M = \$362,981.87 (Mdn = \$337,344.00, SD = \$106,476.61), with the Shapiro–Wilk test confirming a significant deviation from normality, $W = .818$, $p = .001$. The 2021 distribution also demonstrated significant non-normality, albeit to a lesser extent, with M = \$660,182.61 (Mdn = \$619,000.00, SD = \$149,335.68), $W = .895$, $p = .020$. In both instances, the mean surpassed the median, reflecting the impact of high-value housing markets. These findings necessitate consideration when using dwelling value as an outcome variable; however, they do not inherently invalidate linear or quadratic regressions. Ordinary Least Squares (OLS) regression does not require the dependent variable to be normally distributed; normality is primarily pertinent to the residuals in small-sample t and F inference (Schmidt & Finan, 2018).

The most significant deviation from normality was observed among car commuters traveling to Toronto. In 2016, the distribution exhibited a mean of 61,698.91 commuters and a median of only 840 (SD = 264,795.83), whereas in 2021, the mean was 44,789.57 and the median was 895 (SD = 191,454.82). Shapiro–Wilk tests strongly rejected normality for both years, with $W = .243$ and $p < .001$ in each case. The substantial differences between the means and medians, along with the exceptionally large standard deviations, indicate that the commuter distributions are highly right-skewed. This pattern aligns with the notably large Toronto observation compared to other study areas. Consequently, the commuter regressions necessitate considerable caution regarding influential observations, heteroscedasticity, and residual non-normality.

The measure of shelter-cost burden demonstrated a distinct temporal pattern. In 2016, the percentage of owner households allocating more than 30% of their income to shelter was M = 16.64% (Mdn = 16.20%,

SD = 3.40), which significantly deviated from normality, W = .869, p = .006. In contrast, the 2021 distribution was M = 21.37% (Mdn = 22.00%, SD = 4.09) and did not significantly deviate from normality, W = .961, p = .478. Therefore, although the average shelter-cost burden was greater in 2021 compared to 2016, the cross-sectional distribution across study areas in 2021 was more aligned with normality.

**Table 4. Descriptive statistics and Shapiro-Wilk tests.**

| Variable | Mean | Median | SD | Shapiro-Wilk W | p |
|---|---|---|---|---|---|
| Distance to Toronto (km) | 124.74 | 124.9 | 41.69 | 0.959 | 0.435 |
| 2016 Average Dwelling Value ($) | 362981.87 | 337344 | 106476.61 | 0.818 | 0.001 |
| 2021 Average Dwelling Value ($) | 660182.61 | 619000 | 149335.68 | 0.895 | 0.02 |
| 2016 Toronto-bound Car Commuters | 61698.91 | 840 | 264795.83 | 0.243 | < .001 |
| 2021 Toronto-bound Car Commuters | 44789.57 | 895 | 191454.82 | 0.243 | < .001 |
| 2016 Households Spending >30% on Shelter (%) | 16.64 | 16.2 | 3.4 | 0.869 | 0.006 |
| 2021 Households Spending >30% on Shelter (%) | 21.37 | 22 | 4.09 | 0.961 | 0.478 |

### 3.3. *Moran's I spatial correlation testing*

To evaluate whether the observed spatial patterns in housing, affordability, and commuting were geographically clustered rather than randomly distributed, an exploratory spatial autocorrelation analysis was conducted using 2021 Statistics Canada geographic boundary files. Census Metropolitan Area (CMA) polygons were used for observations represented by CMAs, whereas Census Subdivision (CSD) polygons were used for municipalities. Polygon centroids were calculated in the Statistics Canada Lambert Conformal Conic projection (EPSG:3347), and a spatial weights matrix was constructed using a symmetrized four-nearest-neighbor ($k = 4$) approach with row standardization. An inverse-distance spatial weight matrix was also estimated as a robustness check. Global Moran's $I$ statistics, Local Moran's $I$ (LISA), Moran scatterplots, and LISA cluster maps were computed using 999 random permutations to assess statistical significance. Residual Moran's $I$ statistics were calculated for each of the linear and quadratic regression models to determine whether spatial dependence remained after model estimation (Anselin, 1995; Cliff & Ord, 1981).

**Table 5. Summary of spatial autocorrelation analysis method.**

| Item | Specification |
|---|---|
| Geographic matching | CMA polygons for CMA-labelled observations; Census Subdivision polygons for municipal observations. |
| CRS | Statistics Canada Lambert Conformal Conic, EPSG:3347 (metres). |
| Centroids | Geometric polygon centroids calculated in EPSG:3347. |
| Primary weights | Four-nearest-neighbor weights, symmetrized and row-standardized. |
| Robustness weights | Inverse centroid-distance weights among all other study areas, row-standardized. |
| Global inference | Two-sided pseudo-p values from 999 random permutations. |
| Local inference | Conditional randomization with 999 permutations; p < .05 for exploratory cluster classification. |
| Random seed | 20260806 |

The results of the Global Moran's I analysis, as shown in Table 6, reveal that several study variables demonstrated statistically significant positive spatial autocorrelation under the primary four-nearest-neighbor specification. Specifically, the distance from Toronto exhibited significant positive spatial clustering (I = .278, z = 2.80, p = .013), indicating that municipalities in close geographic proximity tended to be situated at similar distances from the metropolitan center. In 2016, average dwelling values also showed significant spatial dependence (I = .212, z = 2.51, p = .036); however, the 2021 estimate, while remaining positive, did not achieve statistical significance (I = .176, p = .151). The most pronounced spatial autocorrelation was identified for the percentage of owner households spending more than 30% of their income on shelter in 2016 (I = .351, z = 3.71, p = .003), suggesting that municipalities with relatively high housing cost burdens were geographically clustered rather than randomly distributed. The corresponding estimate for 2021 was weaker (I = .206) and did not attain conventional statistical significance (p = .083). Conversely, neither average shelter costs nor the volume of Toronto-bound lone car commuters exhibited statistically significant global spatial autocorrelation in either census year, indicating greater spatial heterogeneity for these variables across the study region.

**Table 6. Global Moran Four-nearest-neighbor analysis.**

| Weights | Variable | N | Moran *I* | Expected *I* | Permutation z | Permutation p | Permutations |
|---|---|---|---|---|---|---|---|
| kNN (k=4) | Distance to Toronto (km) | 23 | 0.278 | -0.045 | 2.796 | 0.013 | 999 |
| kNN (k=4) | 2016 average shelter cost | 23 | 0.147 | -0.045 | 1.665 | 0.259 | 999 |
| kNN (k=4) | 2021 average shelter cost | 23 | 0.114 | -0.045 | 1.354 | 0.399 | 999 |
| kNN (k=4) | 2016 average dwelling value | 23 | 0.212 | -0.045 | 2.510 | 0.036 | 999 |
| kNN (k=4) | 2021 average dwelling value | 23 | 0.176 | -0.045 | 1.959 | 0.151 | 999 |
| kNN (k=4) | 2016 Toronto-bound car commuters | 23 | -0.031 | -0.045 | 1.270 | 0.876 | 999 |
| kNN (k=4) | 2021 Toronto-bound car commuters | 23 | -0.029 | -0.045 | 1.429 | 0.905 | 999 |
| kNN (k=4) | 2016 households spending >30% | 23 | 0.351 | -0.045 | 3.705 | 0.003 | 999 |
| kNN (k=4) | 2021 households spending >30% | 23 | 0.206 | -0.045 | 2.247 | 0.083 | 999 |

Source: data analysis

Table 7 assesses the extent to which geographically proximate communities exhibit greater similarity in characteristics compared to those situated at greater distances. The analysis employs Global Moran's I, a metric for spatial autocorrelation, utilizing inverse-distance weighting. This approach ensures that nearby study areas exert a greater influence on each other than those that are more distant. In essence, a positive

Moran's I indicates that communities with similar values are typically located in close proximity, whereas a negative value suggests that neighboring communities are generally dissimilar. A value near zero implies minimal systematic geographic clustering. The permutation p-value determines whether the observed spatial pattern is stronger than what would be expected by chance; at the conventional $\alpha = .05$ level, a p-value less than .05 signifies statistically significant spatial autocorrelation.

Across the 23 study areas, there was generally minimal evidence of significant spatial clustering under the inverse-distance specification. In both 2016 and 2021, average shelter costs, average dwelling values, and Toronto-bound car commuter volumes did not exhibit significant autocorrelation (all $ps > .05$). The primary exception was the percentage of households spending more than 30% of income on shelter costs in 2016, which demonstrated significant positive spatial autocorrelation, Moran's $I = .142$, $p = .004$. This finding indicates that municipalities with similar levels of housing-cost burden were geographically proximate in 2016. Although the corresponding 2021 Moran's I remained positive ($I = .054$), it was not statistically significant, $p = .508$, suggesting that this geographic clustering was no longer apparent by 2021. Overall, the results indicate that most housing and commuting variables were not strongly organized into broad geographic clusters when proximity among all study areas was represented through inverse-distance weighting, with the 2016 shelter-cost burden showing the most pronounced evidence of spatial dependence.

The inverse-distance weighting specification yielded similar overall patterns, albeit with generally weaker evidence of spatial dependence. In this alternative specification, only the 2016 shelter cost burden variable remained statistically significant ($p=0.004$), while the evidence for clustering in dwelling values was diminished. The consistency of the 2016 shelter-burden findings across both weighting schemes indicates that this variable demonstrates relatively robust geographic clustering. In contrast, the spatial structure of dwelling values appears more sensitive to the specification of neighborhood relationships. As alternative spatial weights can affect estimates of spatial dependence, evaluating multiple weighting schemes is an important robustness assessment and is widely recommended in spatial econometric research (Anselin, 2002; Bivand et al., 2013).

**Table 7. Spatial autocorrelation analysis: inverse-distance weighting.**

| Weights | Variable | N | Moran *I* | Expected *I* | Permutation z | Permutation p |
|---|---|---|---|---|---|---|
| Inverse distance | Distance to Toronto (km) | 23 | 0.038 | -0.045 | 2.003 | 0.697 |
| Inverse distance | 2016 average shelter cost | 23 | -0.002 | -0.045 | 0.980 | 0.986 |
| Inverse distance | 2021 average shelter cost | 23 | -0.005 | -0.045 | 0.921 | 0.957 |
| Inverse distance | 2016 average dwelling value | 23 | 0.022 | -0.045 | 1.748 | 0.829 |
| Inverse distance | 2021 average dwelling value | 23 | 0.016 | -0.045 | 1.497 | 0.878 |
| Inverse distance | 2016 Toronto-bound car commuters | 23 | -0.043 | -0.045 | 0.306 | 0.674 |
| Inverse distance | 2021 Toronto-bound car commuters | 23 | -0.043 | -0.045 | 0.322 | 0.683 |
| Inverse distance | 2016 households spending >30% | 23 | 0.142 | -0.045 | 4.675 | 0.004 |
| Inverse distance | 2021 households spending >30% | 23 | 0.054 | -0.045 | 2.482 | 0.508 |

Source: data analysis

Table 8 illustrates the Local Indicators of Spatial Association (LISA) (Anselin, 1995), which determine whether individual communities constitute statistically significant local clusters or spatial outliers in relation to adjacent communities. In contrast to Global Moran's I, which assesses whether a variable exhibits spatial clustering across the entire study region, Local Moran's I examines each community individually. This analysis involves comparing each area's standardized value with the weighted average of its surrounding areas (the spatial lag), and permutation tests are employed to ascertain whether the observed local pattern is unlikely to have occurred by chance. For instance, Toronto exhibited a significant High–High cluster for average shelter costs in both 2016 (I = 2.440, p = .002) and 2021 (I = 2.253, p = .014), signifying that Toronto had relatively high shelter costs and was surrounded by areas with similarly high values. Conversely, St. Catharines–Niagara was identified as Low–High in both years, indicating that its shelter costs were relatively low compared to the higher values of its spatial neighbors.

The clustering categories should be interpreted as a hierarchy of similar clusters versus spatial outliers, rather than a ranking from best to worst. The High–High category denotes a high-value community surrounded by relatively high-value neighbors, while the Low–Low category signifies a low-value community surrounded by relatively low-value neighbours; both represent positive spatial clustering. Conversely, High–Low and Low–High categories identify spatial outliers: a high-value community among relatively low-value neighbours or a low-value community among relatively high-value neighbours, respectively. The term "Not significant" indicates that the observed local configuration was not sufficiently unusual to be statistically distinguished from chance. For instance, the dwelling-value results classify Guelph as High–High in both 2016 (I = .789, p = .015) and 2021 (I = .780, p = .009), whereas St. Catharines–Niagara is categorized as Low–High in both periods, illustrating localized contrasts within the broader regional housing market. It is crucial to note that LISA results describe local geographic patterns rather than causal relationships; a cluster indicates that neighbouring places share unusually similar—or contrasting—values, not that one municipality influences conditions in another (Anselin, 1995).

**Table 8. Local indicators of spatial association of census agglomerations and census metropolitan areas, 2016 and 2021.**

| Z | Centre | Local Moran *I* | Permutation p | Cluster | Standardized value | Spatial lag |
|---|---|---|---|---|---|---|
| 2016 average shelter cost | Toronto CMA | 2.440 | 0.002 | High-High | 2.831 | 0.862 |
| 2016 average shelter cost | Oshawa CMA | 0.069 | 0.925 | Ns | 1.645 | 0.042 |
| 2016 average shelter cost | Hamilton CMA | -0.002 | 0.998 | Ns | 0.885 | -0.002 |
| 2016 average shelter cost | Guelph CMA | 0.754 | 0.075 | Ns | 1.218 | 0.618 |
| 2016 average shelter cost | Brantford CMA | -0.041 | 0.857 | Ns | -0.634 | 0.064 |
| 2016 average shelter cost | Port Hope | 0.025 | 0.727 | Ns | -0.161 | -0.155 |
| 2016 average shelter cost | Centre Wellington | 0.433 | 0.065 | Ns | 0.669 | 0.648 |
| 2016 average shelter cost | Kitchener-Cambridge-Waterloo CMA | 0.113 | 0.664 | Ns | 0.646 | 0.175 |
| 2016 average shelter cost | St. Catharines-Niagara CMA | -0.738 | 0.014 | Low-High | -0.686 | 1.075 |
| 2016 average shelter cost | Barrie CMA | 0.061 | 0.932 | Ns | 1.464 | 0.042 |
| 2016 average shelter cost | Cobourg | -0.029 | 0.904 | Ns | -0.476 | 0.061 |
| 2016 average shelter cost | Kawartha Lakes | -0.092 | 0.790 | Ns | -0.704 | 0.130 |
| 2016 average shelter cost | Orillia | 0.034 | 0.918 | Ns | -0.564 | -0.060 |
| 2016 average shelter cost | Peterborough CMA | 0.043 | 0.793 | Ns | -0.400 | -0.107 |
| 2016 average shelter cost | Woodstock | 0.179 | 0.282 | Ns | -0.453 | -0.395 |
| 2016 average shelter cost | Midland | 0.057 | 0.902 | Ns | -0.908 | -0.063 |
| 2016 average shelter cost | Collingwood | 0.024 | 0.587 | Ns | -0.090 | -0.267 |
| 2016 average shelter cost | Stratford | -0.043 | 0.886 | Ns | -0.634 | 0.067 |
| 2016 average shelter cost | Ingersoll | 0.147 | 0.190 | Ns | -0.225 | -0.653 |
| 2016 average shelter cost | Norfolk | 0.178 | 0.744 | Ns | -1.224 | -0.145 |
| 2016 average shelter cost | Belleville - Quinte West CMA | -0.127 | 0.750 | Ns | -0.838 | 0.152 |
| 2016 average shelter cost | Owen Sound | -0.301 | 0.535 | Ns | -1.060 | 0.284 |
| 2016 average shelter cost | London CMA | 0.191 | 0.197 | Ns | -0.301 | -0.634 |
| 2021 average shelter cost | Toronto CMA | 2.253 | 0.014 | High-High | 2.638 | 0.854 |
| 2021 average shelter cost | Oshawa CMA | 0.025 | 0.976 | Ns | 1.742 | 0.015 |
| 2021 average shelter cost | Hamilton CMA | 0.033 | 0.925 | Ns | 0.836 | 0.039 |
| 2021 average shelter cost | Guelph CMA | 0.675 | 0.069 | Ns | 1.023 | 0.659 |
| 2021 average shelter cost | Brantford CMA | -0.005 | 0.905 | Ns | -0.099 | 0.051 |
| 2021 average shelter cost | Port Hope | 0.019 | 0.766 | Ns | -0.138 | -0.134 |
| 2021 average shelter cost | Centre Wellington | 0.284 | 0.192 | Ns | 0.600 | 0.473 |
| 2021 average shelter cost | Kitchener-Cambridge-Waterloo CMA | 0.121 | 0.585 | Ns | 0.561 | 0.216 |
| 2021 average shelter cost | St. Catharines-Niagara CMA | -0.639 | 0.012 | Low-High | -0.581 | 1.100 |
| 2021 average shelter cost | Barrie CMA | 0.038 | 0.953 | Ns | 1.486 | 0.025 |
| 2021 average shelter cost | Cobourg | 0.004 | 0.988 | Ns | -0.512 | -0.008 |
| 2021 average shelter cost | Kawartha Lakes | -0.036 | 0.787 | Ns | -0.266 | 0.135 |
| 2021 average shelter cost | Orillia | -0.074 | 0.764 | Ns | -0.483 | 0.152 |
| 2021 average shelter cost | Peterborough CMA | 0.026 | 0.909 | Ns | -0.581 | -0.045 |
| 2021 average shelter cost | Woodstock | 0.107 | 0.286 | Ns | -0.266 | -0.403 |
| 2021 average shelter cost | Midland | -0.010 | 0.988 | Ns | -1.063 | 0.010 |
| 2021 average shelter cost | Collingwood | -0.167 | 0.471 | Ns | 0.453 | -0.369 |
| 2021 average shelter cost | Stratford | -0.014 | 0.973 | Ns | -0.802 | 0.018 |
| 2021 average shelter cost | Ingersoll | 0.294 | 0.162 | Ns | -0.433 | -0.678 |
| 2021 average shelter cost | Norfolk | 0.085 | 0.892 | Ns | -1.270 | -0.067 |
| 2021 average shelter cost | Belleville - Quinte West CMA | -0.135 | 0.787 | Ns | -1.053 | 0.128 |
| 2021 average shelter cost | Owen Sound | -0.523 | 0.435 | Ns | -1.418 | 0.369 |
| 2021 average shelter cost | London CMA | 0.259 | 0.152 | Ns | -0.374 | -0.693 |
| 2016 average dwelling value | Toronto CMA | 1.824 | 0.102 | Ns | 3.572 | 0.511 |
| 2016 average dwelling value | Oshawa CMA | 0.286 | 0.464 | Ns | 0.949 | 0.302 |
| 2016 average dwelling value | Hamilton CMA | 0.095 | 0.865 | Ns | 1.176 | 0.080 |
| 2016 average dwelling value | Guelph CMA | 0.789 | 0.015 | High-High | 0.967 | 0.815 |
| 2016 average dwelling value | Brantford CMA | -0.015 | 0.878 | Ns | -0.246 | 0.062 |
| 2016 average dwelling value | Port Hope | 0.014 | 0.733 | Ns | -0.101 | -0.142 |
| 2016 average dwelling value | Centre Wellington | 0.488 | 0.011 | High-High | 0.603 | 0.809 |
| 2016 average dwelling value | Kitchener-Cambridge-Waterloo CMA | 0.056 | 0.682 | Ns | 0.297 | 0.188 |
| 2016 average dwelling value | St. Catharines-Niagara CMA | -0.696 | 0.003 | Low-High | -0.509 | 1.367 |
| 2016 average dwelling value | Barrie CMA | 0.184 | 0.437 | Ns | 0.543 | 0.339 |
| 2016 average dwelling value | Cobourg | 0.023 | 0.786 | Ns | -0.200 | -0.117 |
| 2016 average dwelling value | Kawartha Lakes | -0.001 | 0.988 | Ns | -0.147 | 0.005 |
| 2016 average dwelling value | Orillia | 0.028 | 0.929 | Ns | -0.579 | -0.048 |
| 2016 average dwelling value | Peterborough CMA | 0.028 | 0.798 | Ns | -0.249 | -0.113 |
| 2016 average dwelling value | Woodstock | 0.423 | 0.222 | Ns | -0.926 | -0.456 |
| 2016 average dwelling value | Midland | 0.102 | 0.785 | Ns | -0.853 | -0.120 |
| 2016 average dwelling value | Collingwood | -0.105 | 0.418 | Ns | 0.263 | -0.399 |
| 2016 average dwelling value | Stratford | 0.155 | 0.403 | Ns | -0.448 | -0.347 |
| 2016 average dwelling value | Ingersoll | 0.726 | 0.168 | Ns | -1.120 | -0.648 |
| 2016 average dwelling value | Norfolk | 0.215 | 0.411 | Ns | -0.629 | -0.341 |
| 2016 average dwelling value | Belleville - Quinte West CMA | -0.106 | 0.802 | Ns | -1.065 | 0.100 |
| 2016 average dwelling value | Owen Sound | -0.098 | 0.744 | Ns | -0.707 | 0.139 |
| 2016 average dwelling value | London CMA | 0.460 | 0.110 | Ns | -0.589 | -0.781 |
| 2021 average dwelling value | Toronto CMA | 1.859 | 0.084 | Ns | 3.224 | 0.577 |
| 2021 average dwelling value | Oshawa CMA | 0.169 | 0.632 | Ns | 0.875 | 0.193 |
| 2021 average dwelling value | Hamilton CMA | 0.226 | 0.634 | Ns | 1.135 | 0.199 |
| 2021 average dwelling value | Guelph CMA | 0.780 | 0.009 | High-High | 0.882 | 0.884 |
| 2021 average dwelling value | Brantford CMA | -0.011 | 0.711 | Ns | -0.077 | 0.150 |
| 2021 average dwelling value | Port Hope | -0.015 | 0.551 | Ns | 0.060 | -0.250 |
| 2021 average dwelling value | Centre Wellington | 0.489 | 0.092 | Ns | 0.841 | 0.582 |
| 2021 average dwelling value | Kitchener-Cambridge-Waterloo CMA | 0.108 | 0.594 | Ns | 0.485 | 0.223 |
| 2021 average dwelling value | St. Catharines-Niagara CMA | -0.391 | 0.006 | Low-High | -0.303 | 1.291 |
| 2021 average dwelling value | Barrie CMA | 0.146 | 0.578 | Ns | 0.588 | 0.248 |
| 2021 average dwelling value | Cobourg | 0.049 | 0.721 | Ns | -0.299 | -0.165 |
| 2021 average dwelling value | Kawartha Lakes | -0.007 | 0.956 | Ns | -0.231 | 0.030 |
| 2021 average dwelling value | Orillia | -0.049 | 0.836 | Ns | -0.535 | 0.092 |
| 2021 average dwelling value | Peterborough CMA | 0.051 | 0.667 | Ns | -0.282 | -0.181 |
| 2021 average dwelling value | Woodstock | 0.281 | 0.265 | Ns | -0.665 | -0.422 |
| 2021 average dwelling value | Midland | 0.119 | 0.768 | Ns | -0.795 | -0.150 |
| 2021 average dwelling value | Collingwood | -0.444 | 0.247 | Ns | 0.807 | -0.550 |
| 2021 average dwelling value | Stratford | 0.135 | 0.654 | Ns | -0.778 | -0.173 |
| 2021 average dwelling value | Ingersoll | 0.658 | 0.188 | Ns | -1.017 | -0.647 |
| 2021 average dwelling value | Norfolk | 0.144 | 0.568 | Ns | -0.635 | -0.226 |
| 2021 average dwelling value | Belleville - Quinte West CMA | -0.116 | 0.841 | Ns | -1.313 | 0.089 |
| 2021 average dwelling value | Owen Sound | -0.525 | 0.415 | Ns | -1.458 | 0.360 |
| 2021 average dwelling value | London CMA | 0.393 | 0.109 | Ns | -0.508 | -0.774 |
| 2016 Toronto-bound car commuters | Toronto CMA | -0.841 | 0.953 | Ns | 4.684 | -0.179 |
| 2016 Toronto-bound car commuters | Oshawa CMA | -0.036 | 0.231 | Ns | -0.062 | 0.585 |
| 2016 Toronto-bound car commuters | Hamilton CMA | 0.012 | 0.692 | Ns | -0.052 | -0.223 |
| 2016 Toronto-bound car commuters | Guelph CMA | -0.132 | 0.034 | Low-High | -0.211 | 0.624 |
| 2016 Toronto-bound car commuters | Brantford CMA | 0.045 | 0.888 | Ns | -0.231 | -0.194 |
| 2016 Toronto-bound car commuters | Port Hope | 0.047 | 0.723 | Ns | -0.236 | -0.200 |
| 2016 Toronto-bound car commuters | Centre Wellington | -0.147 | 0.027 | Low-High | -0.235 | 0.624 |
| 2016 Toronto-bound car commuters | Kitchener-Cambridge-Waterloo CMA | 0.041 | 0.854 | Ns | -0.203 | -0.201 |
| 2016 Toronto-bound car commuters | St. Catharines-Niagara CMA | -0.233 | 0.005 | Low-High | -0.222 | 1.047 |
| 2016 Toronto-bound car commuters | Barrie CMA | -0.125 | 0.218 | Ns | -0.167 | 0.747 |
| 2016 Toronto-bound car commuters | Cobourg | 0.046 | 0.729 | Ns | -0.237 | -0.192 |
| 2016 Toronto-bound car commuters | Kawartha Lakes | 0.044 | 0.766 | Ns | -0.227 | -0.192 |
| 2016 Toronto-bound car commuters | Orillia | 0.051 | 0.660 | Ns | -0.237 | -0.217 |
| 2016 Toronto-bound car commuters | Peterborough CMA | 0.047 | 0.742 | Ns | -0.234 | -0.200 |
| 2016 Toronto-bound car commuters | Woodstock | 0.054 | 0.476 | Ns | -0.237 | -0.230 |
| 2016 Toronto-bound car commuters | Midland | 0.052 | 0.638 | Ns | -0.237 | -0.220 |
| 2016 Toronto-bound car commuters | Collingwood | 0.052 | 0.618 | Ns | -0.237 | -0.220 |
| 2016 Toronto-bound car commuters | Stratford | 0.054 | 0.500 | Ns | -0.238 | -0.229 |
| 2016 Toronto-bound car commuters | Ingersoll | 0.056 | 0.264 | Ns | -0.238 | -0.236 |
| 2016 Toronto-bound car commuters | Norfolk | 0.047 | 0.763 | Ns | -0.237 | -0.198 |
| 2016 Toronto-bound car commuters | Belleville - Quinte West CMA | 0.046 | 0.749 | Ns | -0.238 | -0.192 |
| 2016 Toronto-bound car commuters | Owen Sound | 0.052 | 0.589 | Ns | -0.237 | -0.219 |
| 2016 Toronto-bound car commuters | London CMA | 0.055 | 0.196 | Ns | -0.230 | -0.238 |
| 2021 Toronto-bound car commuters | Toronto CMA | -0.820 | 0.961 | Ns | 4.685 | -0.175 |
| 2021 Toronto-bound car commuters | Oshawa CMA | -0.031 | 0.219 | Ns | -0.052 | 0.586 |
| 2021 Toronto-bound car commuters | Hamilton CMA | 0.016 | 0.678 | Ns | -0.073 | -0.222 |
| 2021 Toronto-bound car commuters | Guelph CMA | -0.130 | 0.039 | Low-High | -0.210 | 0.621 |
| 2021 Toronto-bound car commuters | Brantford CMA | 0.045 | 0.860 | Ns | -0.230 | -0.197 |
| 2021 Toronto-bound car commuters | Port Hope | 0.047 | 0.800 | Ns | -0.236 | -0.197 |
| 2021 Toronto-bound car commuters | Centre Wellington | -0.145 | 0.030 | Low-High | -0.234 | 0.621 |
| 2021 Toronto-bound car commuters | Kitchener-Cambridge-Waterloo CMA | 0.041 | 0.803 | Ns | -0.201 | -0.204 |
| 2021 Toronto-bound car commuters | St. Catharines-Niagara CMA | -0.231 | 0.007 | Low-High | -0.221 | 1.043 |
| 2021 Toronto-bound car commuters | Barrie CMA | -0.118 | 0.223 | Ns | -0.158 | 0.747 |
| 2021 Toronto-bound car commuters | Cobourg | 0.045 | 0.812 | Ns | -0.238 | -0.190 |
| 2021 Toronto-bound car commuters | Kawartha Lakes | 0.043 | 0.856 | Ns | -0.225 | -0.190 |
| 2021 Toronto-bound car commuters | Orillia | 0.051 | 0.663 | Ns | -0.237 | -0.214 |
| 2021 Toronto-bound car commuters | Peterborough CMA | 0.046 | 0.810 | Ns | -0.233 | -0.198 |
| 2021 Toronto-bound car commuters | Woodstock | 0.055 | 0.456 | Ns | -0.237 | -0.230 |
| 2021 Toronto-bound car commuters | Midland | 0.052 | 0.637 | Ns | -0.238 | -0.217 |
| 2021 Toronto-bound car commuters | Collingwood | 0.052 | 0.615 | Ns | -0.237 | -0.218 |
| 2021 Toronto-bound car commuters | Stratford | 0.055 | 0.493 | Ns | -0.239 | -0.229 |
| 2021 Toronto-bound car commuters | Ingersoll | 0.057 | 0.228 | Ns | -0.239 | -0.237 |
| 2021 Toronto-bound car commuters | Norfolk | 0.048 | 0.695 | Ns | -0.238 | -0.202 |
| 2021 Toronto-bound car commuters | Belleville - Quinte West CMA | 0.045 | 0.828 | Ns | -0.237 | -0.190 |
| 2021 Toronto-bound car commuters | Owen Sound | 0.052 | 0.584 | Ns | -0.238 | -0.217 |
| 2021 Toronto-bound car commuters | London CMA | 0.056 | 0.197 | Ns | -0.233 | -0.238 |
| 2016 households spending >30% | Toronto CMA | 0.382 | 0.740 | Ns | 3.023 | 0.126 |
| 2016 households spending >30% | Oshawa CMA | 0.169 | 0.286 | Ns | 0.409 | 0.414 |
| 2016 households spending >30% | Hamilton CMA | -0.022 | 0.239 | Ns | 0.048 | -0.457 |
| 2016 households spending >30% | Guelph CMA | -0.029 | 0.667 | Ns | -0.162 | 0.179 |
| 2016 households spending >30% | Brantford CMA | 0.242 | 0.236 | Ns | -0.553 | -0.437 |
| 2016 households spending >30% | Port Hope | -0.002 | 0.936 | Ns | 0.048 | -0.036 |
| 2016 households spending >30% | Centre Wellington | -0.090 | 0.775 | Ns | -0.763 | 0.118 |
| 2016 households spending >30% | Kitchener-Cambridge-Waterloo CMA | 0.427 | 0.118 | Ns | -0.733 | -0.583 |
| 2016 households spending >30% | St. Catharines-Niagara CMA | 0.028 | 0.195 | Ns | 0.048 | 0.589 |
| 2016 households spending >30% | Barrie CMA | 1.276 | 0.002 | High-High | 1.100 | 1.160 |
| 2016 households spending >30% | Cobourg | 0.010 | 0.869 | Ns | -0.132 | -0.079 |
| 2016 households spending >30% | Kawartha Lakes | 0.080 | 0.661 | Ns | 0.319 | 0.251 |
| 2016 households spending >30% | Orillia | 0.706 | 0.025 | High-High | 0.679 | 1.040 |
| 2016 households spending >30% | Peterborough CMA | 0.000 | 0.999 | Ns | -0.132 | 0.000 |
| 2016 households spending >30% | Woodstock | 0.974 | 0.032 | Low-Low | -1.244 | -0.783 |
| 2016 households spending >30% | Midland | 0.504 | 0.076 | Ns | 0.619 | 0.814 |
| 2016 households spending >30% | Collingwood | 0.931 | 0.362 | Ns | 2.121 | 0.439 |
| 2016 households spending >30% | Stratford | 0.781 | 0.027 | Low-Low | -0.823 | -0.949 |
| 2016 households spending >30% | Ingersoll | 1.077 | 0.053 | Ns | -1.274 | -0.845 |
| 2016 households spending >30% | Norfolk | 0.437 | 0.066 | Ns | -0.583 | -0.751 |
| 2016 households spending >30% | Belleville - Quinte West CMA | -0.031 | 0.927 | Ns | -0.643 | 0.048 |
| 2016 households spending >30% | Owen Sound | -0.494 | 0.082 | Ns | -0.643 | 0.769 |
| 2016 households spending >30% | London CMA | 0.719 | 0.043 | Low-Low | -0.733 | -0.981 |
| 2021 households spending >30% | Toronto CMA | 0.291 | 0.734 | Ns | 2.281 | 0.127 |
| 2021 households spending >30% | Oshawa CMA | 0.128 | 0.529 | Ns | 0.532 | 0.241 |
| 2021 households spending >30% | Hamilton CMA | -0.161 | 0.299 | Ns | 0.407 | -0.396 |
| 2021 households spending >30% | Guelph CMA | 0.042 | 0.504 | Ns | 0.157 | 0.266 |
| 2021 households spending >30% | Brantford CMA | 0.021 | 0.812 | Ns | -0.217 | -0.096 |
| 2021 households spending >30% | Port Hope | 0.004 | 0.936 | Ns | -0.117 | -0.037 |
| 2021 households spending >30% | Centre Wellington | -0.269 | 0.511 | Ns | -1.116 | 0.241 |
| 2021 households spending >30% | Kitchener-Cambridge-Waterloo CMA | -0.053 | 0.389 | Ns | 0.157 | -0.334 |
| 2021 households spending >30% | St. Catharines-Niagara CMA | 0.054 | 0.174 | Ns | 0.083 | 0.657 |
| 2021 households spending >30% | Barrie CMA | 0.959 | 0.006 | High-High | 0.982 | 0.977 |
| 2021 households spending >30% | Cobourg | -0.008 | 0.965 | Ns | 0.357 | -0.024 |
| 2021 households spending >30% | Kawartha Lakes | -0.273 | 0.321 | Ns | -0.567 | 0.482 |
| 2021 households spending >30% | Orillia | 0.650 | 0.296 | Ns | 1.331 | 0.488 |
| 2021 households spending >30% | Peterborough CMA | -0.018 | 0.824 | Ns | 0.182 | -0.097 |
| 2021 households spending >30% | Woodstock | -0.049 | 0.019 | High-Low | 0.058 | -0.858 |
| 2021 households spending >30% | Midland | 0.214 | 0.104 | Ns | 0.257 | 0.832 |
| 2021 households spending >30% | Collingwood | 0.738 | 0.219 | Ns | 1.281 | 0.576 |
| 2021 households spending >30% | Stratford | 0.842 | 0.106 | Ns | -1.291 | -0.652 |
| 2021 households spending >30% | Ingersoll | 1.581 | 0.137 | Ns | -2.590 | -0.611 |
| 2021 households spending >30% | Norfolk | 0.608 | 0.321 | Ns | -1.441 | -0.422 |
| 2021 households spending >30% | Belleville - Quinte West CMA | -0.165 | 0.617 | Ns | -0.692 | 0.239 |
| 2021 households spending >30% | Owen Sound | -0.094 | 0.465 | Ns | -0.267 | 0.351 |
| 2021 households spending >30% | London CMA | -0.306 | 0.003 | High-Low | 0.232 | -1.316 |

Note: Ns: not significant.
Source: Data analysis

Table 9 reports Global Moran's *I* tests applied to the residuals of the linear and quadratic regression models. This is an important diagnostic because ordinary least-squares regression assumes that, after the modeled relationship has been accounted for, the remaining errors are reasonably independent. In spatial data, however, nearby communities may have similar unexplained outcomes. Moran's *I* provides a formal test of whether these residuals are geographically patterned rather than randomly distributed (Moran, 1950; Anselin, 1995). Here, spatial relationships were defined using a four-nearest-neighbours (kNN, $k = 4$) weighting scheme, meaning that each study area was linked to its four geographically closest study areas. The permutation test then assessed whether the observed residual Moran's *I* was more extreme than would be expected under spatial randomness. A significant positive Moran's *I* indicates that nearby communities tend to have residuals in the same direction—that is, the model systematically under- or overpredicts groups of neighbouring communities. A nonsignificant result indicates no statistically detectable residual spatial clustering.

The results provide strong support for the spatial independence of the commuting and dwelling-value models. Residual autocorrelation was nonsignificant for both linear and quadratic commuter models in 2016 and 2021 (all $p$s $\geq .146$) and for all dwelling-value models (all $p$s $\geq .506$). This is particularly important because the quadratic models explained substantial variation in commuter volumes (2016: $R^2 = .767$; 2021: $R^2 = .766$) and dwelling values (2016: $R^2 = .860$; 2021: $R^2 = .805$) without leaving statistically significant geographic clustering in their residuals. Thus, although these outcomes are geographically structured, there is little evidence that the fitted models systematically missed additional spatial patterns among neighbouring communities.

A different pattern emerged for owner shelter-cost burden. In 2016, substantial positive residual autocorrelation remained in both the distance models—linear, Moran's $I = .445$, $p = .002$, and quadratic, $I = .518$, $p = .001$—and the dwelling-value models—linear, $I = .576$, $p = .001$, and quadratic, $I = .604$, $p = .001$.

In practical terms, neighbouring communities tended to have similar levels of shelter-cost burden that were not fully explained by either distance from Toronto or average dwelling value. This suggests that other geographically concentrated factors—such as local incomes, housing tenure, housing supply, mortgage conditions, or regional socioeconomic characteristics—may contribute to affordability outcomes. The pattern weakened in 2021 but did not disappear: residual autocorrelation was significant for the dwelling-value linear model ($I$ = .235, $p$ = .048), the distance quadratic model ($I$ = .273, $p$ = .021), and the dwelling-value quadratic model ($I$ = .242, $p$ = .039), whereas the distance linear model was not significant ($I$ = .202, $p$ = .093).

Overall, the residual Moran's $I$ analysis strengthens confidence in the commuting and dwelling-value regression results because their residuals showed no statistically significant spatial dependence under the four-nearest-neighbours specification. Conversely, the persistent clustering of shelter-burden residuals indicates that these models do not fully capture the spatial processes underlying housing affordability. Accordingly, coefficients from the shelter-burden OLS models should be interpreted more cautiously, and spatial regression approaches—such as spatial-error or spatial-lag models—would provide useful extensions. Moran's $I$ and related spatial diagnostics identify spatial association rather than causation; significant residual clustering demonstrates that unexplained values are geographically related, but does not establish that conditions in one municipality cause those in neighbouring municipalities (Moran, 1950; Anselin, 1995).

Residual Moran's $I$ provides a diagnostic assessment of whether spatial dependence remains in the regression errors after estimation of an ordinary least squares model (Anselin, 2002; Anselin & Bera, 1998). For the commuter and average dwelling value models, the residual spatial autocorrelation was generally weak and statistically non-significant, indicating that the estimated linear and quadratic regressions adequately captured the principal spatial gradients present in the data. This finding supports the assumption of spatial independence underlying the regression analyses and suggests that the reported coefficient estimates and statistical inferences are unlikely to be substantially biased by omitted spatial processes.

**Table 9. Residual Moran's *I* – four-nearest-neighbors, kNN (k=4) weighting.**

| Model | Moran $I$ | Expected I | Permutation z | Permutation p | $R^2$ | Adjusted $R^2$ | N |
|---|---|---|---|---|---|---|---|
| 2016 commuters vs distance - Linear | -0.141 | -0.045 | -1.125 | 0.146 | 0.331 | 0.299 | 23 |
| 2016 dwelling value vs distance - Linear | -0.027 | -0.045 | 0.159 | 0.838 | 0.722 | 0.709 | 23 |
| 2016 shelter burden vs distance - Linear | 0.445 | -0.045 | 4.493 | 0.002 | 0.270 | 0.235 | 23 |
| 2016 shelter burden vs dwelling value - Linear | 0.576 | -0.045 | 5.579 | 0.001 | 0.463 | 0.437 | 23 |
| 2016 commuters vs distance - Quadratic | -0.138 | -0.045 | -0.887 | 0.270 | 0.767 | 0.744 | 23 |
| 2016 dwelling value vs distance - Quadratic | 0.020 | -0.045 | 0.519 | 0.899 | 0.860 | 0.846 | 23 |
| 2016 shelter burden vs distance - Quadratic | 0.518 | -0.045 | 5.257 | 0.001 | 0.345 | 0.279 | 23 |
| 2016 shelter burden vs dwelling value - Quadratic | 0.604 | -0.045 | 5.955 | 0.001 | 0.482 | 0.430 | 23 |
| 2021 commuters vs distance - Linear | -0.139 | -0.045 | -1.099 | 0.155 | 0.331 | 0.299 | 23 |
| 2021 dwelling value vs distance - Linear | -0.057 | -0.045 | -0.127 | 0.684 | 0.754 | 0.742 | 23 |
| 2021 shelter burden vs distance - Linear | 0.202 | -0.045 | 2.118 | 0.093 | 0.236 | 0.199 | 23 |
| 2021 shelter burden vs dwelling value - Linear | 0.235 | -0.045 | 2.395 | 0.048 | 0.365 | 0.335 | 23 |
| 2021 commuters vs distance - Quadratic | -0.136 | -0.045 | -0.868 | 0.279 | 0.766 | 0.743 | 23 |
| 2021 dwelling value vs distance - Quadratic | -0.090 | -0.045 | -0.414 | 0.506 | 0.805 | 0.786 | 23 |
| 2021 shelter burden vs distance - Quadratic | 0.273 | -0.045 | 2.760 | 0.021 | 0.290 | 0.219 | 23 |
| 2021 shelter burden vs dwelling value - Quadratic | 0.242 | -0.045 | 2.466 | 0.039 | 0.369 | 0.306 | 23 |

In contrast, several regression models predicting the proportion of owner households spending more than 30% of their income on shelter retained statistically significant positive residual Moran's *I* values, particularly for the 2016 analyses. These findings suggest that although distance from Toronto and average dwelling value explain a substantial proportion of the observed variation in shelter cost burden, additional spatially structured factors remain unaccounted for within the regression models. Such factors may include regional labor-market conditions, municipal planning policies, local housing supply constraints, transportation accessibility, or neighborhood socioeconomic composition, all of which may exhibit geographic clustering (LeSage & Pace, 2009). Consequently, future investigations of housing affordability would benefit from estimating spatial errors or spatial lag models, or by employing spatially robust standard errors to explicitly account for any remaining spatial dependence.

Overall, spatial diagnostic analyses reinforce the principal findings of the study rather than contradicting them. The observed geographic clustering of dwelling values and housing-cost burden is consistent with the existence of regional housing-market structures across Southern Ontario, while the absence of significant spatial autocorrelation in commuter volumes suggests that commuting patterns are influenced primarily by distance-related relationships identified in the regression analyses rather than by residual geographic clustering. The generally non-significant residual Moran's *I* statistics for the commuter and dwelling-value regressions further support the robustness of the estimated linear and quadratic models. Consequently, spatial diagnostics strengthen confidence in the principal conclusions of this study while also identifying shelter-cost burden as the outcome for which future spatial econometric modelling would be most informative.

### 3.4. *Lone car commuters*

Table 10 presents the results of quadratic least squares regression models examining the association between distance from Toronto and the number of lone car commuters in 2016 and 2021. In both years, the overall regression models were statistically significant, 2016: $F(2, 20) = 32.93$, $p < .001$, $R^2 = .767$, adjusted $R^2 = .744$; 2021: $F(2, 20) = 32.76$, $p < .001$, $R^2 = .766$, adjusted $R^2 = .743$. The unstandardized intercept (constant) decreased from 1,346,568.12 in 2016 (SE = 162,645.19, 95% CI [1,007,296.20, 1,685,840.05]) to 973,166.83 in 2021 (SE = 117,826.53, 95% CI [727,384.99, 1,218,948.67]). The linear distance coefficient remained negative and statistically significant in both years but decreased in absolute magnitude from $B = -20{,}207.71$ (SE = 2,791.14, $t = -7.24$, $p < .001$) in 2016 to $B = -14{,}598.58$ (SE = 2,022.01, $t = -7.22$, $p < .001$) in 2021. Similarly, the quadratic coefficient was positive and statistically significant in both years, declining from $B = 71.76$ (SE = 11.73, $t = 6.12$, $p < .001$) in 2016 to $B = 51.83$ (SE = 8.50, $t = 6.10$, $p < .001$) in 2021. The models exhibited nearly identical goodness-of-fit statistics across years, with similar $R^2$ values, adjusted $R^2$ values, F-statistics, and residual degrees of freedom. The model selection criteria were lower in 2021 (AIC = 596.30; BIC = 599.71) than in 2016 (AIC = 611.13; BIC = 614.54), and both the residual standard error and in-sample RMSE were smaller in 2021. All analyses were based on the same sample of 23 census metropolitan areas and communities.

**Table 10. Quadratic regression: Lone car commuters to Toronto by distance from Toronto.**

| Statistic | 2016 | 2021 |
|---|---|---|
| Constant, B | 1,346,568.122 | 973,166.833 |
| Constant, SE | 162,645.193 | 117,826.532 |
| Constant, t | 8.279 | 8.259 |
| Constant, p | 0.000 | 0.000 |
| Constant, 95% CI lower | 1,007,296.195 | 727,384.995 |
| Constant, 95% CI upper | 1,685,840.049 | 1,218,948.672 |
| Distance (km), B | -20,207.707 | -14,598.584 |
| Distance (km), SE | 2,791.135 | 2,022.007 |
| Distance (km), t | -7.240 | -7.220 |
| Distance (km), p | 0.000 | 0.000 |
| Distance (km), 95% CI lower | -26,029.912 | -18,816.416 |
| Distance (km), 95% CI upper | -14,385.503 | -10,380.751 |
| Distance², B | 71.757 | 51.831 |
| Distance², SE | 11.729 | 8.497 |
| Distance², t | 6.118 | 6.100 |
| Distance², p | 0.000 | 0.000 |
| Distance², 95% CI lower | 47.291 | 34.106 |
| Distance², 95% CI upper | 96.223 | 69.555 |
| Model F | 32.926 | 32.759 |
| Numerator df | 2.000 | 2.000 |
| Denominator df | 20.000 | 20.000 |
| Model p | 0.000 | 0.000 |
| R² | 0.767 | 0.766 |
| Adjusted R² | 0.744 | 0.743 |
| Residual standard error | 134,043.482 | 97,106.336 |
| RMSE (in-sample) | 124,996.192 | 90,552.125 |
| AIC | 611.129 | 596.300 |
| BIC | 614.535 | 599.707 |
| Residual sum of squares | 359,353,103,700.739 | 188,592,808,122.761 |
| N | 23 | 23 |

Source: data analysis

Table 11 presents the ordinary least squares regression models examining the association between distance from Toronto and the number of lone car commuters in 2016 and 2021.The linear OLS regression results indicate a statistically significant negative association between distance from Toronto and Toronto-bound lone car commuter volumes in both 2016 and 2021. In 2016, distance significantly predicted commuter volume, $B = -3,654.53$, $SE = 1,133.60$, $t(21) = -3.22$, $p = .004$, 95% CI [−6,011.97, −1,297.08]. The overall model was significant, $F(1, 21) = 10.39$, $p = .004$, and explained 33.1% of the variation in commuter volumes ($R^2 = .331$, adjusted $R^2 = .299$). A comparable relationship was observed in 2021, with distance again significantly and negatively associated with commuter volume, $B = -2,642.19$, $SE = 819.64$, $t(21) = -3.22$, $p = .004$, 95% CI [−4,346.73, −937.64]. The 2021 model was also significant, $F(1, 21) = 10.39$, $p = .004$, and accounted for the same proportion of variance ($R^2 = .331$, adjusted $R^2 = .299$).

Comparison of the two years reveals considerable stability in the strength of the distance–commuting relationship, despite lower absolute commuter volumes in 2021. Although the magnitude of the distance coefficient declined from −3,654.53 in 2016 to −2,642.19 in 2021, both models produced virtually identical standardized statistical evidence and explanatory power. The lower residual standard error in 2021 (160,277.31 versus 221,669.25 in 2016) is also consistent with the lower overall scale of commuter volumes in that year. Collectively, the results indicate that communities located farther from Toronto consistently generated fewer Toronto-bound lone car commuters, while the unchanged $R^2$ suggests that the relative importance of distance in explaining the geographic distribution of commuting remained stable between 2016 and 2021.

**Table 11. OLS regression: Lone car commuters to Toronto by distance from Toronto.**

| Statistic | 2016 | 2021 |
|---|---|---|
| Constant, B | 517,577.249 | 374,384.901 |
| Constant, SE | 148,771.203 | 107,568.584 |
| Constant, t | 3.479 | 3.480 |
| Constant, p | 0.002 | 0.002 |
| Constant, 95% CI lower | 208,190.596 | 150,683.784 |
| Constant, 95% CI upper | 826,963.902 | 598,086.018 |
| Distance (km), B | -3,654.526 | -2,642.185 |
| Distance (km), SE | 1,133.597 | 819.644 |
| Distance (km), t | -3.224 | -3.224 |
| Distance (km), p | 0.004 | 0.004 |
| Distance (km), 95% CI lower | -6,011.971 | -4,346.728 |
| Distance (km), 95% CI upper | -1,297.082 | -937.641 |
| Model F | 10.393 | 10.391 |
| Numerator df | 1.000 | 1.000 |
| Denominator df | 21.000 | 21.000 |
| Model p | 0.004 | 0.004 |
| $R^2$ | 0.331 | 0.331 |
| Adjusted $R^2$ | 0.299 | 0.299 |
| Residual standard error | 221,669.252 | 160,277.306 |
| RMSE (in-sample) | 211,812.304 | 153,150.269 |
| AIC | 633.390 | 618.473 |
| BIC | 635.661 | 620.744 |
| Residual sum of squares | 1,031,882,403,462.020 | 539,465,110,572.586 |
| N | 23 | 23 |

Source: data analysis

Figure 2 illustrates the relationship between distance from Toronto and the number of lone car commuters for 2016 and 2021, together with the fitted quadratic regression curves and their associated 95% confidence intervals. In both years, the observed data exhibited a pronounced nonlinear pattern, characterized by relatively high commuter counts among communities closest to Toronto, followed by a marked decline as distance increased. The fitted quadratic curves closely followed this pattern, with the 2016 curve generally lying above the corresponding 2021 curve across most of the observed distance range, reflecting the lower commuter counts recorded in 2021. The greatest concentration of observations occurred between approximately 90 and 160 km from Toronto, where commuter volumes were comparatively low and the fitted curves converged before gradually increasing toward the most distant communities. The confidence intervals are narrowest across the central portion of the distance distribution, where observations are most densely clustered, and widen at shorter and longer distances, where relatively few communities are represented, indicating greater uncertainty in the fitted values at the extremes of the observed range. The fitted quadratic regressions reached their minimum predicted commuter counts at approximately 140.8 km from Toronto in both years, indicating that the location of the turning point remained essentially unchanged between the two census years despite differences in the estimated intercept and regression coefficients. Overall, the figure provides a graphical representation of the quadratic relationships estimated by the regression models and illustrates the distribution of the observed commuter counts relative to the fitted trajectories for both census years.

**Figure 2. Plot of linear and quadratic regressions of lone car commuters to Toronto by distance from Toronto.**

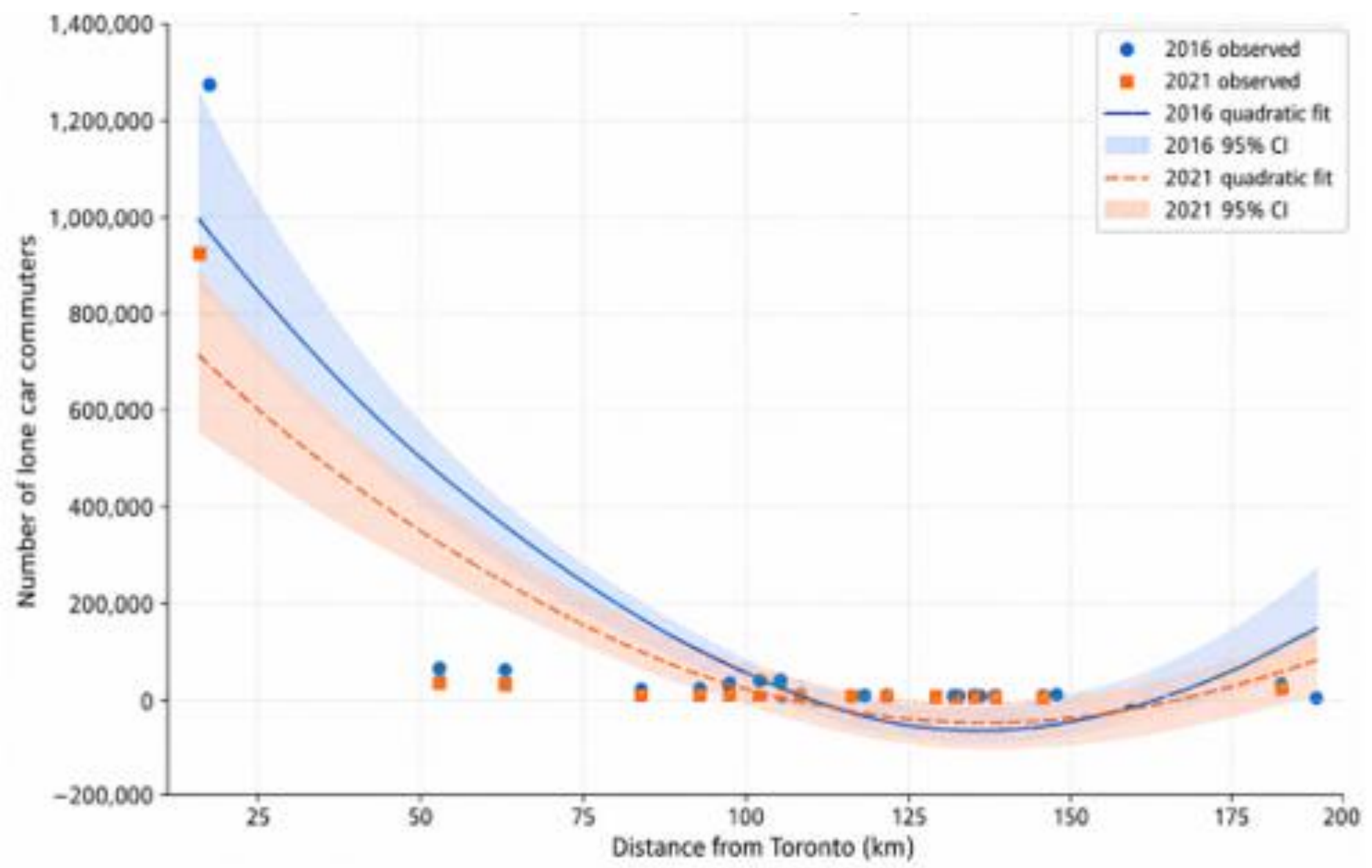

### 3.5. *Average dwelling value*

Table 12 summarizes the linear regression models relating the average dwelling value to the distance from Toronto for 2016 and 2021. In both years, the overall regression model was statistically significant, 2016: $F(1, 21) = 10.39$, $p = .004$, $R^2 = .331$, adjusted $R^2 = .299$; 2021: $F(1, 21) = 10.39$, $p = .004$, $R^2 = .331$, adjusted $R^2 = .299$. The intercept (constant) decreased from **517,577.25** in 2016 (SE = 148,771.20, 95% CI [208,190.60, 826,963.90]) to 374,384.90 in 2021 (SE = 107,568.58, 95% CI [150,683.78, 598,086.02]), although the corresponding *t-statistics* ($t = 3.48$) and significance levels ($p = .002$) were identical across both years. The linear distance coefficient remained negative and statistically significant in both models, decreasing in absolute magnitude from $B = −3,654.53$ (SE = 1,133.60, $t = −3.22$, $p = .004$, 95% CI [−6,011.97, −1,297.08]) in 2016 to $B = −2,642.19$ (SE = 819.64, $t = −3.22$, $p = .004$, 95% CI [−4,346.73, −937.64]) in 2021. The two linear models exhibited identical goodness-of-fit statistics, with $R^2 = .331$ and adjusted $R^2 = .299$ in both years, as well as the same numerator and denominator degrees of freedom. Measures of model error, however, differed between years, with lower residual standard error (160,277.31 vs. 221,669.25), lower in-sample RMSE (153,150.27 vs. 211,812.30), lower residual sum of squares ($5.39 \times 10^{11}$ vs. $1.03 \times 10^{12}$), and smaller information criteria (AIC = 618.47, BIC = 620.74) in 2021 compared with 2016 (AIC = 633.39, BIC = 635.66). Both regression models were estimated using the same sample of 23 census metropolitan areas and communities.

**Table 12. Linear regressions: 2016 and 2021 average dwelling value versus distance to Toronto.**

| Statistic | 2016 | 2021 |
|---|---|---|
| Constant, B | 517,577.249 | 374,384.901 |
| Constant, SE | 148,771.203 | 107,568.584 |
| Constant, t | 3.479 | 3.480 |
| Constant, p | 0.002 | 0.002 |
| Constant, 95% CI lower | 208,190.596 | 150,683.784 |
| Constant, 95% CI upper | 826,963.902 | 598,086.018 |
| Distance (km), B | -3,654.526 | -2,642.185 |
| Distance (km), SE | 1,133.597 | 819.644 |
| Distance (km), t | -3.224 | -3.224 |
| Distance (km), p | 0.004 | 0.004 |
| Distance (km), 95% CI lower | -6,011.971 | -4,346.728 |
| Distance (km), 95% CI upper | -1,297.082 | -937.641 |
| Model F | 10.393 | 10.391 |
| Numerator df | 1.000 | 1.000 |
| Denominator df | 21.000 | 21.000 |
| Model p | 0.004 | 0.004 |
| $R^2$ | 0.331 | 0.331 |
| Adjusted $R^2$ | 0.299 | 0.299 |
| Residual standard error | 221,669.252 | 160,277.306 |
| RMSE (in-sample) | 211,812.304 | 153,150.269 |
| AIC | 633.390 | 618.473 |
| BIC | 635.661 | 620.744 |
| Residual sum of squares | 1,031,882,403,462.020 | 539,465,110,572.586 |
| N | 23 | 23 |

Source: data analysis

Table 13 presents quadratic regression models relating the average dwelling value to distance from Toronto for 2016 and 2021. In both years, the overall models were statistically significant, 2016: $F(2, 20) = 32.93$, $p < .001$, $R^2 = .767$, adjusted $R^2 = .744$; 2021: $F(2, 20) = 32.76$, $p < .001$, $R^2 = .766$, adjusted $R^2 = .743$. The intercept (constant) decreased from 1,346,568.12 in 2016 (SE = 162,645.19, 95% CI [1,007,296.20, 1,685,840.05]) to 973,166.83 in 2021 (SE = 117,826.53, 95% CI [727,384.99, 1,218,948.67]), while remaining statistically significant in both models ($p < .001$). The linear distance coefficient was negative and statistically significant in both years, declining in absolute magnitude from $B = -20{,}207.71$ (SE = 2,791.14, $t = -7.24$, $p < .001$) in 2016 to $B = -14{,}598.58$ (SE = 2,022.01, $t = -7.22$, $p < .001$) in 2021. The quadratic Distance² coefficient was positive and statistically significant in both years, decreasing from $B = 71.76$ (SE = 11.73, $t = 6.12$, $p < .001$) in 2016 to $B = 51.83$ (SE = 8.50, $t = 6.10$, $p < .001$) in 2021. The two quadratic models exhibited nearly identical goodness-of-fit statistics, with $R^2$ values of .767 and .766 and adjusted $R^2$ values of .744 and .743 for 2016 and 2021, respectively. Measures of model error were lower in 2021 than in 2016, including the residual standard error (97,106.34 vs. 134,043.48), in-sample RMSE (90,552.13 vs. 124,996.19), residual sum of squares ($1.89 \times 10^{11}$ vs. $3.59 \times 10^{11}$), and information criteria (AIC = 596.30, BIC = 599.71 versus AIC = 611.13, BIC = 614.54).

Compared with the corresponding linear regression models, the quadratic models had substantially larger absolute regression coefficients because they partitioned the relationship into both linear and quadratic components, and demonstrated a markedly improved model fit, with $R^2$ increasing from .331 to approximately .767 in both years, adjusted $R^2$ increasing from .299 to approximately .744, and lower residual standard errors, RMSE values, AIC values, BIC values, and residual sums of squares. Therefore, the inclusion of the quadratic term altered the estimated regression parameters while substantially increasing the proportion of variation accounted for by the fitted models, although both the linear and quadratic specifications were estimated using the same sample of 23 census metropolitan areas and communities.

**Table 13. Quadratic regressions: 2016 and 2021 average dwelling values versus distance to Toronto.**

| Statistic | 2016 | 2021 |
|---|---|---|
| Constant, B | 1,346,568.122 | 973,166.833 |
| Constant, SE | 162,645.193 | 117,826.532 |
| Constant, t | 8.279 | 8.259 |
| Constant, p | 0.000 | 0.000 |
| Constant, 95% CI lower | 1,007,296.195 | 727,384.995 |
| Constant, 95% CI upper | 1,685,840.049 | 1,218,948.672 |
| Distance (km), B | -20,207.707 | -14,598.584 |
| Distance (km), SE | 2,791.135 | 2,022.007 |
| Distance (km), t | -7.240 | -7.220 |
| Distance (km), p | 0.000 | 0.000 |
| Distance (km), 95% CI lower | -26,029.912 | -18,816.416 |
| Distance (km), 95% CI upper | -14,385.503 | -10,380.751 |

| Distance², B | 71.757 | 51.831 |
|---|---|---|
| Distance², SE | 11.729 | 8.497 |
| Distance², t | 6.118 | 6.100 |
| Distance², p | 0.000 | 0.000 |
| Distance², 95% CI lower | 47.291 | 34.106 |
| Distance², 95% CI upper | 96.223 | 69.555 |
| Model F | 32.926 | 32.759 |
| Numerator df | 2.000 | 2.000 |
| Denominator df | 20.000 | 20.000 |
| Model p | 0.000 | 0.000 |
| $R^2$ | 0.767 | 0.766 |
| Adjusted $R^2$ | 0.744 | 0.743 |
| Residual standard error | 134,043.482 | 97,106.336 |
| RMSE (in-sample) | 124,996.192 | 90,552.125 |
| AIC | 611.129 | 596.300 |
| BIC | 614.535 | 599.707 |
| Residual sum of squares | 359,353,103,700.739 | 188,592,808,122.761 |
| N | 23 | 23 |

Source: data analysis

Figure 3 illustrates the relationship between distance from Toronto and the average dwelling value for 2016 and 2021, together with the fitted linear and quadratic regression models. Observed dwelling values are shown for each census metropolitan area and community, while the fitted regression lines summarize the overall trend in housing values as the distance from Toronto increases. In both years, the average dwelling values were highest among communities located closest to Toronto, and generally declined with increasing distance. The 2021 observations consistently lie above the corresponding 2016 observations across most of the study area, reflecting higher average dwelling values in 2021.

**Figure 3. Average dwelling values versus distance to Toronto.**

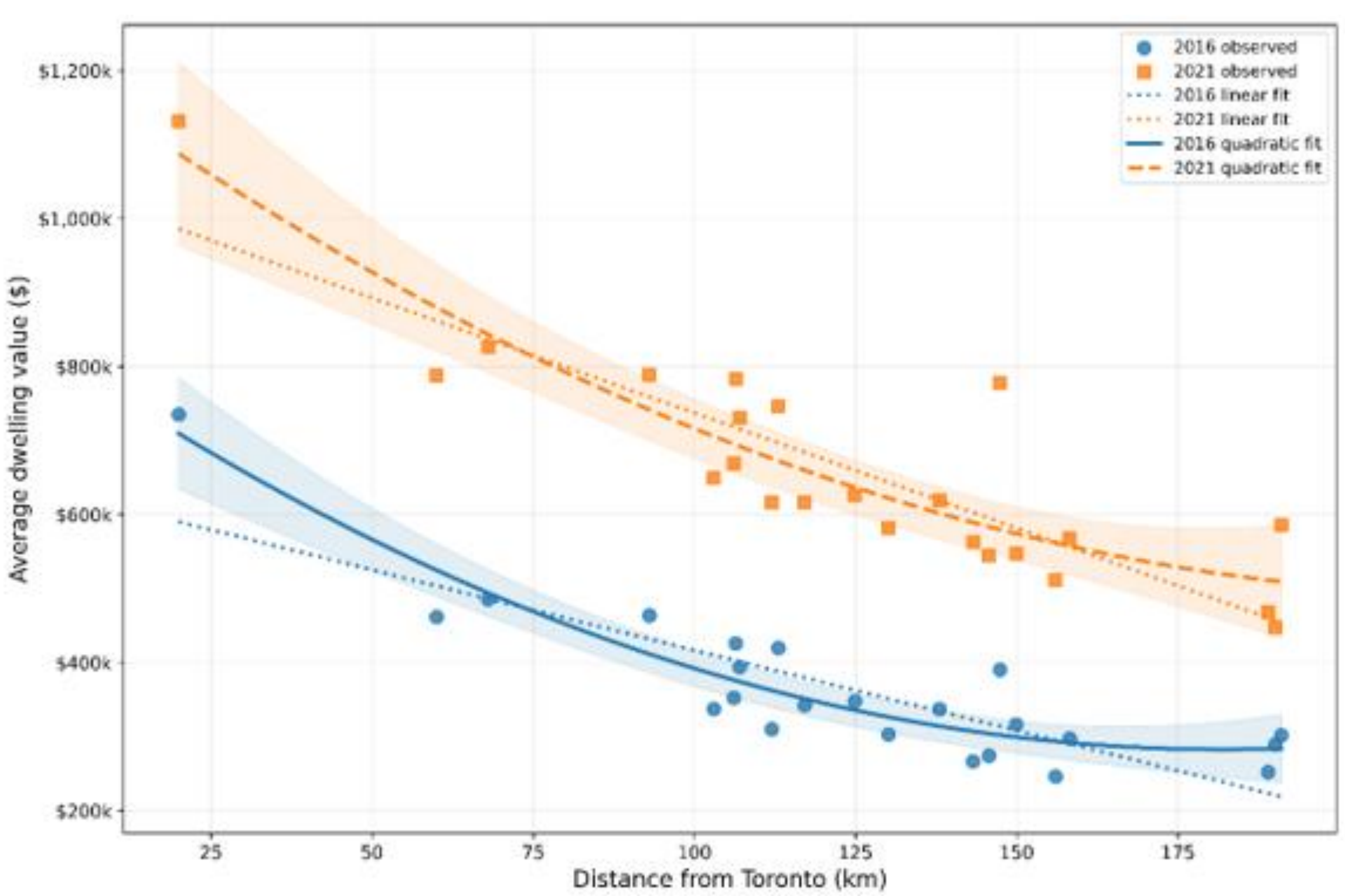

The linear regression models, represented by dotted lines, indicate a consistent decrease in average dwelling value as the distance from Toronto increases. In contrast, the quadratic models, depicted by solid and dashed lines, reveal a more pronounced nonlinear relationship. In both years analyzed, the quadratic curves exhibited a steep decline in dwelling values over shorter distances from Toronto, followed by a progressively flatter gradient at greater distances. This curvature aligns with the statistically significant quadratic terms identified in the regression analysis, illustrating the diminishing rate of decline in predicted dwelling values as distance increases. The fitted quadratic curves consistently remained above the corresponding linear regression lines across much of the observed distance range, gradually converging at larger distances. This suggests that the quadratic specification offers a distinct representation of the relationship compared to the assumption of a constant linear rate of change. Although both modeling approaches depict declining dwelling values with increasing distance, the quadratic models more accurately reflect the observed data distribution, particularly between approximately 90 and 160 km from Toronto, where most communities are situated. The shaded regions surrounding the quadratic regression curves denote 95% confidence intervals for the mean predicted dwelling values. These intervals are narrowest in the central portion of the distance distribution, where observations are most densely concentrated, and widen toward both the shortest and longest distances from Toronto due to the relatively few communities represented at the extremes of the study area. Consequently, the precision of the predicted values is greatest within the middle range of the observed distances and decreases near the data boundaries.

The estimated quadratic models indicated that the minimum predicted dwelling values were located at approximately 182.2 km from Toronto in 2016 and 227.5 km in 2021. In 2016, this minimum was situated at the upper end of the observed distance range, whereas the estimated minimum for 2021 extended beyond the most distant community included in the dataset, which was 191 km. Consequently, the 2021 curve continues to decline across the observed distances, failing to reach its theoretical minimum within the available observations. This variation in the minima's locations reflects changes in the estimated linear and quadratic regression coefficients between the two census years, demonstrating that the fitted 2021 relationship maintains a downward slope throughout the observed study area.

Regression models reveal a general decline in housing values with increasing distance from Toronto, although the estimated rate of decline varies depending on the modelling approach. Linear regression models indicate that the average dwelling value decreased by approximately \$3,655 per kilometre from Toronto in 2016 and \$2,642 per kilometre in 2021. These coefficients reflect the average change in dwelling value per additional kilometre from Toronto, assuming a constant rate of decline across the study area. In contrast, quadratic regression models suggest that the rate of decline is not constant but varies with distance. In 2016, the initial slope of the fitted relationship was −\$20,208 per kilometre, while in 2021, it was −\$14,599 per kilometre. The positive and statistically significant quadratic coefficients for both years (2016: B = 71.76; 2021: B = 51.83) indicate that the magnitude of the negative slope diminishes with increasing distance from Toronto, flattening the fitted curves at greater distances. Consequently, the reduction in predicted dwelling value per kilometre is most pronounced among communities closer to Toronto and decreases as distance increases, rather than remaining constant across the study area. This nonlinear pattern suggests that the impact of distance on average dwelling value is strongest within the inner portion of the study region and diminishes toward more distant communities.

### 3.6. *Shelter cost burden – households spending more than 30% of income on shelter*

Table 14 presents the quadratic regression models that associate the percentage of owner households spending over 30% of their income on shelter with the distance from Toronto for the years 2016 and 2021. In both instances, the overall regression models demonstrated statistical significance: 2016: $F(2, 20) = 5.26$, $p = .015$, $R^2 = .345$, adjusted $R^2 = .279$; 2021: $F(2, 20) = 4.09$, $p = .032$, $R^2 = .290$, adjusted $R^2 = .219$. The intercept increased from 26.35 in 2016 (SE = 3.51, 95% CI [19.04, 33.67]) to 31.84 in 2021 (SE = 4.39, 95% CI [22.69, 41.00]), suggesting a higher predicted percentage of owner households surpassing the 30% shelter-cost threshold at the reference point of the fitted model. The linear distance coefficient was negative in both years, with a slight decrease from B = −0.131 (SE = 0.060, t = −2.17, p = .042) in 2016 to B = −0.138 (SE = 0.075, t = −1.83, p = .082) in 2021. The linear term was statistically significant in 2016 but not in 2021. The quadratic coefficient remained positive in both years, though it was small in magnitude and not statistically significant in either model (2016: B = 0.000383, p = .145; 2021: B = 0.000392, p = .230), indicating limited evidence for curvature after accounting for the linear distance effect.

A comparison of the two models shows that the 2016 regression explained a somewhat larger proportion of the variation in shelter cost burden than the 2021 model ($R^2 = .345$ versus .290; adjusted $R^2$ = .279 versus .219). Consistent with this, measures of model error increased between census years, with the residual standard error rising from 2.89 to 3.62, the in-sample RMSE increasing from 2.69 to 3.37, and the residual sum of squares increasing from 166.95 to 261.70. Similarly, both information criteria were larger in 2021 (AIC = 127.20, BIC = 130.61) than in 2016 (AIC = 116.86, BIC = 120.27). Although the overall regression remained statistically significant in both years, the results indicate that the 2021 model exhibited a modest reduction in explanatory power and precision relative to the corresponding 2016 model, while being estimated from the same sample of 23 census metropolitan areas and communities.

**Table 14. Quadratic regression: Percentage of households spending 30% or more of income versus distance to Toronto, 2016 and 2021.**

| Statistic | 2016 | 2021 |
|---|---|---|
| Constant, B | 26.352 | 31.843 |
| Constant, SE | 3.506 | 4.389 |
| Constant, t | 7.517 | 7.255 |
| Constant, p | 0.000 | 0.000 |
| Constant, 95% CI lower | 19.039 | 22.687 |
| Constant, 95% CI upper | 33.665 | 40.999 |
| Distance (km), B | -0.131 | -0.138 |
| Distance (km), SE | 0.060 | 0.075 |
| Distance (km), t | -2.173 | -1.833 |
| Distance (km), p | 0.042 | 0.082 |
| Distance (km), 95% CI lower | -0.256 | -0.295 |
| Distance (km), 95% CI upper | -0.005 | 0.019 |
| $Distance^2$, B | 0.000 | 0.000 |
| $Distance^2$, SE | 0.000 | 0.000 |
| $Distance^2$, t | 1.515 | 1.238 |
| $Distance^2$, p | 0.145 | 0.230 |
| $Distance^2$, 95% CI lower | 0.000 | 0.000 |
| $Distance^2$, 95% CI upper | 0.001 | 0.001 |
| Model F | 5.263 | 4.087 |
| Numerator df | 2.000 | 2.000 |
| Denominator df | 20.000 | 20.000 |
| Model p | 0.015 | 0.032 |
| $R^2$ | 0.345 | 0.290 |
| Adjusted $R^2$ | 0.279 | 0.219 |
| Residual standard error | 2.889 | 3.617 |
| RMSE (in-sample) | 2.694 | 3.373 |
| AIC | 116.862 | 127.200 |
| BIC | 120.269 | 130.607 |
| Residual sum of squares | 166.953 | 261.700 |
| N | 23 | 23 |

Source: data analysis

Table 15 displays the ordinary least squares models examining the relationship between average housing value and distance for the years 2016 and 2021. The analysis reveals a robust and statistically significant inverse relationship between distance from Toronto and average dwelling value in both census years. In 2016, distance exhibited a negative association with dwelling value, $B = -2{,}170.18$, $SE = 293.84$, $t(21) = -7.39$, $p < .001$, with a 95% confidence interval of [−2,781.25, −1,559.10]. The model was significant overall, $F(1, 21) = 54.55$, $p < .001$, accounting for 72.2% of the variance in average dwelling values ($R^2 = .722$, adjusted $R^2 = .709$). In 2021, the negative spatial gradient intensified, $B = -3{,}110.07$, $SE = 387.81$, $t(21) = -8.02$, $p < .001$, 95% CI [−3,916.57, −2,303.57], with the model explaining 75.4% of the variance, $F(1, 21) = 64.31$, $p < .001$, $R^2 = .754$, adjusted $R^2 = .742$.

A comparison of the two years indicates that the housing-value gradient centered on Toronto not only persisted but also became more pronounced from 2016 to 2021. The estimated decrease in dwelling value per additional kilometre of distance increased from approximately $2,170 in 2016 to $3,110 in 2021, while the explained variance rose modestly from 72.2% to 75.4%. Concurrently, the significantly higher intercept in 2021 aligns with the general increase in housing values throughout the study region. These findings collectively suggest that widespread housing appreciation did not negate the spatial premium associated with proximity to Toronto; instead, housing values rose across the region, and the absolute value differential related to distance expanded. These results offer strong empirical evidence for a persistent and increasingly pronounced spatial housing-value gradient centered on Toronto.

**Table 15. OLS regression: Percentage of households spending 30% or more of income versus distance to Toronto, 2016 and 2021.**

| Statistic | 2016 | 2021 |
|---|---|---|
| Constant, B | 633,697.017 | 1,048,143.246 |
| Constant, SE | 38,563.192 | 50,895.740 |
| Constant, t | 16.433 | 20.594 |
| Constant, p | < .001 | < .001 |
| Constant, 95% CI lower | 553,500.469 | 942,299.760 |
| Constant, 95% CI upper | 713893.566 | 1153986.732 |
| Distance (km), B | -2,170.175 | -3,110.067 |
| Distance (km), SE | 293.841 | 387.812 |
| Distance (km), t | -7.386 | -8.020 |
| Distance (km), p | < .001 | < .001 |
| Distance (km), 95% CI lower | -2,781.251 | -3,916.567 |
| Distance (km), 95% CI upper | -1559.098 | -2303.568 |
| Model F | 54.546 | 64.313 |
| Numerator df | 1 | 1 |
| Denominator df | 21 | 21 |
| Model p | < .001 | < .001 |
| $R^2$ | 0.722 | 0.754 |
| Adjusted $R^2$ | 0.709 | 0.742 |
| Residual standard error | 57,459.198 | 75,834.708 |
| RMSE (in-sample) | 54904.165 | 72462.572 |
| AIC | 571.285 | 584.049 |
| BIC | 573.556 | 586.320 |
| Residual sum of squares | 69,332,748,332.098 | 120,768,960,265.087 |

| N | 23 | 23 |
|---|---|---|

Source: data analysis

Figure 4 depicts the correlation between the distance from Toronto and the percentage of owner households allocating more than 30% of their income to shelter in the years 2016 and 2021. This is presented alongside the fitted quadratic regression curves and their corresponding 95% confidence intervals. In both census years, the data reveal a general decrease in the proportion of owner households facing high shelter cost burdens as the distance from Toronto increases. The quadratic regression curves suggest that this decrease is notably steep in communities nearest to Toronto, subsequently leveling off at greater distances, with the rate of change diminishing beyond approximately 150–175 km from the metropolitan center.

**Figure 4. Percentage of households spending more than 30% of income versus distance from Toronto.**

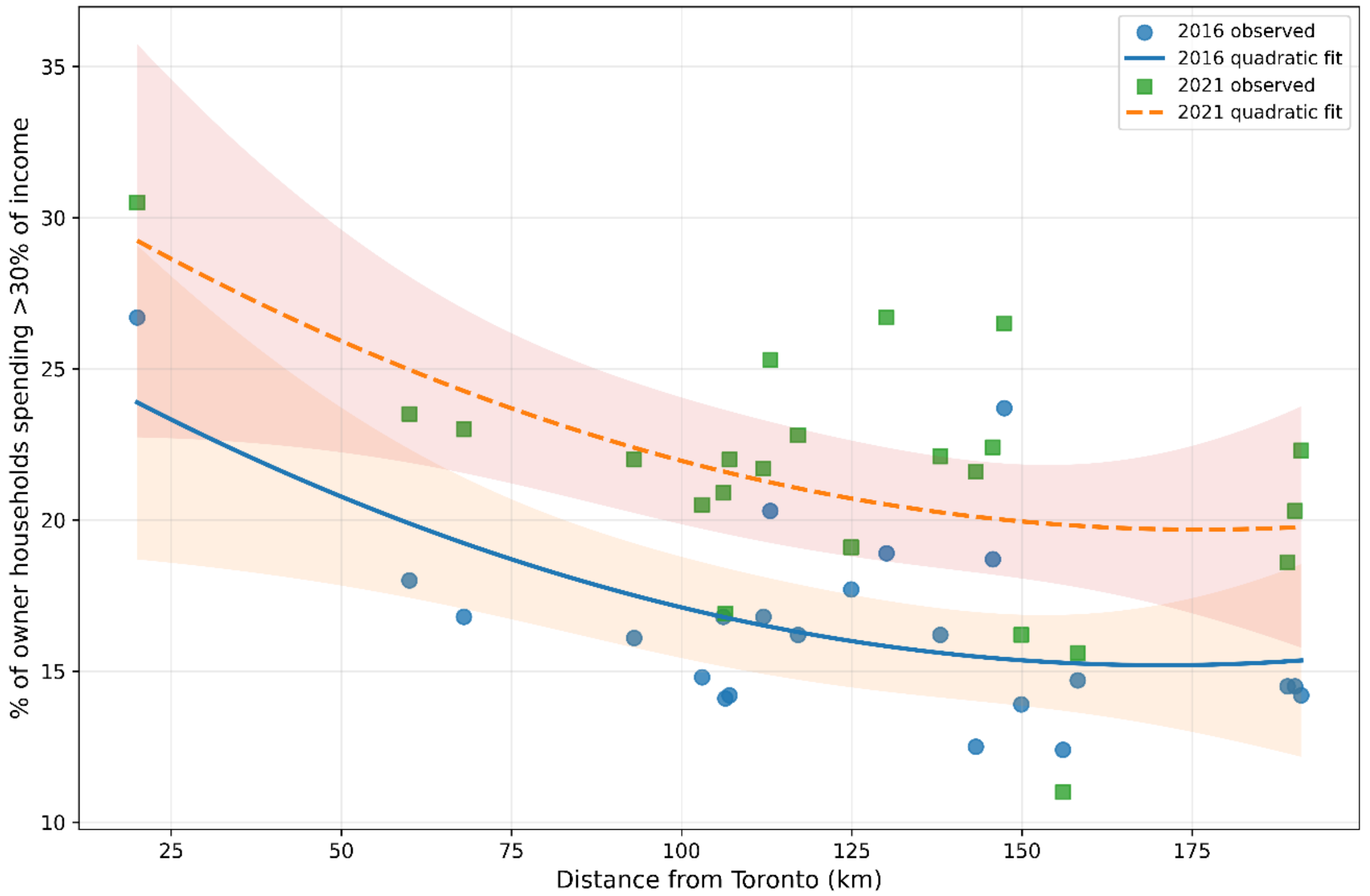

A comparative analysis of the two census years, 2016 and 2021, demonstrated a consistent increase in the observed percentages across much of the study area. In most communities, the 2021 data points (represented by green squares) are situated above the corresponding 2016 data points (depicted by blue circles). This positioning suggests that a greater proportion of owner-households allocated more than 30% of their income to housing in 2021 compared to 2016, at similar distances from Toronto. This upward trend is also evident in the fitted quadratic regression curves, with the 2021 curve consistently positioned above the 2016 curve across nearly the entire observed distance range. Although both curves display similar nonlinear characteristics, the vertical separation between them indicates that the increase in the prevalence of shelter cost burden was widespread across communities throughout the region, rather than being limited to areas immediately adjacent to Toronto.

The analysis of the fitted curves indicates that the relative difference between the two census years was consistent across short, intermediate, and long commuting distances. Although the proportion of owner households surpassing the 30% shelter cost threshold generally decreased with increasing distance in both years, the 2021 curve consistently projected higher percentages than the 2016 curve. This pattern is characterized by significant overlap in the confidence intervals, particularly at the shortest and longest distances from Toronto, where fewer observations are available, indicating greater uncertainty in the fitted values at these extremes. Conversely, the confidence intervals were narrower in the central portion of the study area, where communities are more densely represented, resulting in more precise estimated mean responses.

The quadratic models estimated for the years 2016 and 2021 reached their minimum predicted values at approximately 170.7 km and 176.2 km from Toronto, respectively. The proximity of these turning points suggests that the locations where the fitted curves attained their lowest percentages experienced only minor changes between the census years. While the overall form of the relationship remained similar, the upward shift of the 2021 regression curve compared to the 2016 curve indicates a consistently higher proportion of owner households spending more than 30% of their income on shelter throughout the study region in 2021. This occurred while maintaining a generally similar spatial pattern concerning the distance from Toronto.

### 3.7. *Shelter cost burden versus average dwelling value*

Table 16 displays the outcomes of distinct linear regression models that investigate the association between average dwelling value and the percentage of owner households allocating more than 30% of their income to shelter for the years 2016 and 2021. The linear model for 2016 demonstrated statistical significance ($F(1, 21) = 18.08$, $p < .001$, $R^2 = .463$, adjusted $R^2 = .437$). There was a positive correlation between average dwelling value and the percentage of owner households spending over 30% of their income on shelter, with $B = 0.00002174$, $SE = 0.00000511$, $t = 4.25$, $p < .001$. This coefficient, when expressed per $100,000, indicates an estimated increase of approximately 2.17 percentage points. Similarly, the 2021 linear model was statistically significant, $F(1, 21) = 12.09$, $p = .002$, $R^2 = .365$, adjusted $R^2 = .335$. The coefficient for dwelling value was positive, $B = 0.00001657$, $SE = 0.00000477$, $t = 3.48$, $p = .002$, corresponding to an increase of approximately 1.66 percentage points per $100,000.

**Table 16. Linear regression: Percentage of households spending more than 30% of household income on shelter costs versus average dwelling value in 2016 and 2021.**

| Statistic | 2016 | 2021 |
|---|---|---|
| Constant, B | 8.748 | 10.431 |
| Constant, SE | 1.931 | 3.222 |
| Constant, t | 4.531 | 3.237 |
| Constant, p | 0.000 | 0.004 |
| Constant, 95% CI lower | 4.732 | 3.730 |
| Constant, 95% CI upper | 12.763 | 17.132 |
| Average dwelling value ($), B | 0.000022 | 0.000017 |
| Average dwelling value ($), SE | 0.000005 | 0.000005 |
| Average dwelling value ($), t | 4.252 | 3.477 |
| Average dwelling value ($), p | 0.000 | 0.002 |
| Average dwelling value ($), 95% CI lower | 0.000011 | 0.000007 |
| Average dwelling value ($), 95% CI upper | 0.000032 | 0.000026 |
| Model F | 18.082 | 12.089 |
| Numerator df | 1 | 1 |
| Denominator df | 21 | 21 |
| Model p | 0.000 | 0.002 |
| $R^2$ | 0.463 | 0.365 |
| Adjusted $R^2$ | 0.437 | 0.335 |
| Residual standard error | 2.553 | 3.338 |
| RMSE (in-sample) | 2.440 | 3.189 |
| AIC | 110.301 | 122.625 |
| BIC | 112.572 | 124.896 |
| Residual sum of squares | 136.921 | 233.976 |
| N | 23 | 23 |

Source: data analysis

The quadratic models demonstrated statistical significance overall for both 2016, $F(2, 20) = 9.29$, $p = .001$, $R^2 = .482$, adjusted $R^2 = .430$, and 2021, $F(2, 20) = 5.84$, $p = .010$, $R^2 = .369$, adjusted $R^2 = .306$. Nonetheless, neither the linear nor the squared dwelling-value terms achieved individual statistical significance within either quadratic model. The quadratic specification resulted in only marginal increases in explained variance compared to the corresponding linear models and exhibited higher AIC and BIC values.

Figure 5 depicts the correlation between average dwelling value and the percentage of owner households allocating more than 30% of their income to shelter in the years 2016 and 2021. The data for each community are shown alongside the fitted linear and quadratic regression models, including their 95% confidence intervals. In both census years, a positive correlation is apparent, with communities possessing higher average dwelling values typically showing a larger proportion of owner households whose shelter costs surpassed the 30% affordability threshold. Despite notable variability among individual communities, the general trend suggests that increases in average dwelling value are associated with elevated levels of housing cost burden among owning households.

**Figure 5. Percentage of households spending more than 30% of income versus average dwelling value**

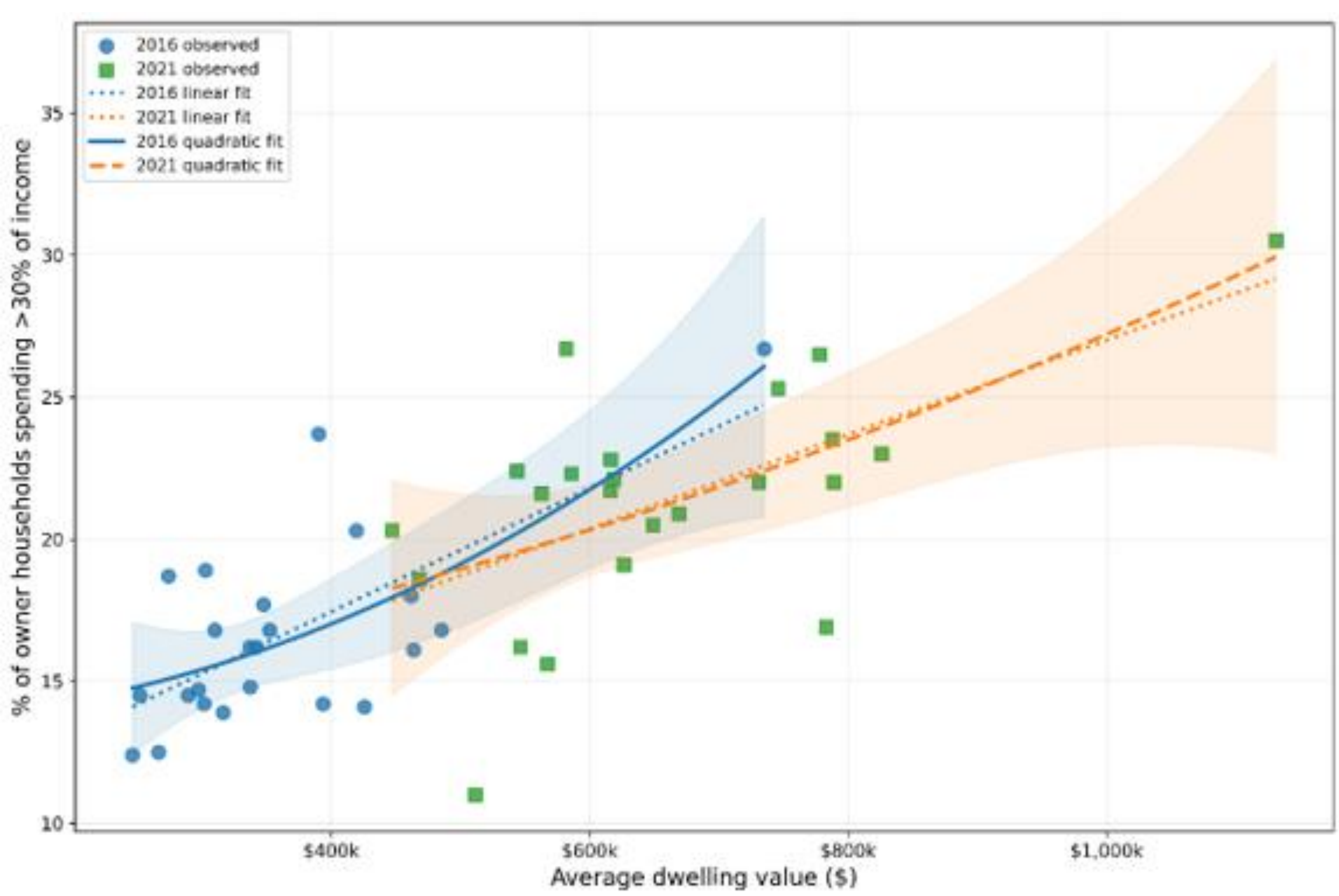

A comparison of linear and quadratic regression models reveals that both approaches capture a similar positive relationship, yet they differ in their depiction of the rate of change. Linear models assume a constant increase in the percentage of cost-burdened households with each incremental rise in average dwelling value, resulting in straight-line relationships across the observed values. Conversely, quadratic models permit the slope to vary within the housing market. For the 2016 data, the quadratic curve demonstrates a slight upward curvature, indicating that the predicted proportion of owner households exceeding the 30% shelter-cost threshold increases more rapidly in communities with higher average dwelling values than a strictly linear relationship would suggest. Nonetheless, the difference between the linear and quadratic fits was relatively minor across the observed range, aligning with regression results where the quadratic term was not statistically significant. Similarly, the 2021 quadratic curve deviates only slightly from the linear fit, indicating that accounting for nonlinear curvature offers limited improvement over the simpler linear specification. The considerable overlap between the linear and quadratic predictions across most dwelling values further illustrates that both models describe broadly similar data patterns.

A comparative analysis of the relationships in 2016 and 2021 indicates that the observations from 2021 are generally elevated compared to those from 2016. This suggests that communities with similar average dwelling values exhibit a higher percentage of owner households spending over 30% of their income on shelter in 2021. This upward shift is notably evident among communities with average dwelling values ranging from approximately $500,000 to $800,000, where many 2021 observations surpass the corresponding 2016 values. The fitted regression lines also illustrate that the anticipated proportion of cost-burdened households remained consistently higher in 2021 across much of the observed dwelling value distribution. At the highest dwelling values, the 2021 models continue to predict an increasing shelter cost burden, whereas the 2016 models show a more gradual rise over the same range of housing values. These differences are reflected in the higher intercepts and distinct coefficient estimates observed in the 2021 regression models, while maintaining the overall positive association between dwelling value and housing affordability pressures.

The confidence intervals for the fitted regression curves are narrowest in the central portion of the dwelling-value distribution, where observations are most densely concentrated. These intervals widen toward both the lower and upper extremes of the housing market due to the relatively sparse distribution of communities in these ranges. As a result, predictions were estimated with greater precision for communities with intermediate dwelling values compared to those at the lowest and highest ends of the observed distribution.

The figure illustrates a consistent positive correlation between average dwelling value and the proportion of owner households allocating more than 30% of their income to shelter costs. Communities with lower average dwelling values typically show a smaller proportion of owner households facing high shelter cost burdens, whereas those with higher-valued housing markets tend to have a larger percentage of households surpassing the affordability threshold. Although the quadratic models allow for slight deviations from linearity, the data predominantly indicate a monotonic and positive relationship in both census years. In conclusion, the figure indicates that higher average dwelling values are generally linked to a greater prevalence of owner housing affordability challenges. Furthermore, the entire relationship shifted upward between 2016 and 2021, suggesting that by 2021, a larger share of owner households exceeded the 30% shelter cost benchmark across nearly the full range of dwelling values.

## 5. Discussion

The primary objective of this study was to examine whether the substantial rise in residential property values from 2016 to 2021 was associated with changes in the spatial distribution of housing affordability and Toronto-oriented automobile commuting across southern Ontario. The findings collectively demonstrate that average dwelling values, owner shelter cost burden, and commuter volumes exhibited distinct spatial gradients relative to their distance from Toronto, although the strength and form of these relationships varied significantly. In all analyses, average dwelling values consistently displayed a more pronounced spatial structure than monthly shelter costs, indicating that housing purchase prices were more closely linked to residential location decisions than current housing expenditures. Regression analyses revealed that average dwelling values systematically decreased with increasing distance from Toronto, with quadratic models providing a significantly better fit to the observed data than linear models. This marked improvement in model fit supports the notion that metropolitan housing markets are characterized by nonlinear spatial gradients rather than constant rates of price change (Dubin & Sung, 1987). Dwelling values decreased most sharply in communities closest to Toronto, then gradually leveled off in more distant areas, suggesting that the accessibility premium associated with proximity to Toronto diminished with distance. This finding is consistent with urban spatial-equilibrium theory, which posits that accessibility to major employment centers is reflected in residential property values through bid-rent mechanisms (So et al., 2001). The persistence of this nonlinear gradient in both census years indicates that Toronto remained the dominant economic center within the regional housing market, despite the significant appreciation observed throughout southern Ontario.

Although housing values have significantly increased across nearly all communities, the estimated spatial gradient has remained stable. The housing value curve shifted upward from 2016 to 2021, reflecting widespread appreciation rather than a fundamental alteration in the relationship between distance and property values. Many peripheral communities experienced appreciation rates approaching or exceeding 100%, indicating that the influence of Toronto's housing market extended beyond the metropolitan boundary. This trend aligns with recent findings that housing demand expanded into suburban and exurban markets during the COVID-19 period, as households sought more affordable housing while maintaining access to metropolitan employment opportunities (Delventhal et al., 2022; Davis et al., 2024; Van Nieuwerburgh, 2023). Rather than eliminating the traditional distance gradient, rapid appreciation appears to have elevated housing prices throughout the regional market.

The analysis of the cost burden associated with owner-occupied shelters revealed a weaker spatial correlation compared to average dwelling values. Although the proportion of owner households spending more than 30% of their income on shelter generally decreased with increasing distance from Toronto, the models exhibited limited explanatory power, and evidence for nonlinear curvature was minimal. Importantly, the entire relationship shifted upward between 2016 and 2021, indicating a decline in housing affordability for owners throughout the study region. This widespread increase in shelter cost burden corresponds with the substantial rise in housing values during the study period and supports previous research distinguishing between housing purchase prices and current shelter expenditures (Van Nieuwerburgh, 2023). Existing homeowners may experience relatively stable monthly housing costs due to historical mortgage arrangements, whereas the rapid appreciation of dwelling values significantly increases financial barriers for prospective buyers. Consequently, average dwelling value appears to reflect changes more accurately in housing market accessibility than current shelter expenditures.

The observed positive correlation between average dwelling value and owner shelter cost burden further supports this interpretation. Communities with higher average dwelling values exhibited a greater proportion of owner households exceeding the conventional 30% affordability threshold, indicating that elevated housing markets are generally associated with increased affordability pressures. The relatively minor improvement from the quadratic specifications suggests that this relationship is predominantly linear across the observed range of dwelling values. These findings imply that, regardless of geographic location, increases in residential property values are accompanied by heightened housing affordability challenges, aligning with broader evidence of declining housing affordability in rapidly appreciating metropolitan regions (Van Nieuwerburgh, 2023).

The analysis of Toronto-oriented automobile commuting revealed a distinctly different spatial pattern. Distance from Toronto accounted for a substantial proportion of the variation in commuter counts, with quadratic models explaining approximately three-quarters of the observed variability. Commuter volumes declined sharply with increasing distance before stabilizing among more distant communities, consistent with classical expectations that commuting becomes progressively less attractive as travel costs rise (Roberts and Taylor, 2017; So et al., 2001). Importantly, widespread dwelling-value appreciation did not translate into a corresponding increase in Toronto-bound commuting: the magnitude of housing appreciation explained only 4.5% of the variation in commuter change and was not a significant predictor of that change ($R^2 = .045$, $p = .331$). This divergence suggests that the geographic expansion of housing demand extended beyond the spatial expansion of conventional automobile commuting, and that residential decentralization did not necessarily produce proportional growth in daily travel to Toronto.

The disparity observed between housing appreciation and commuter volume is notably significant within the context of the COVID-19 pandemic. Current research in urban economics suggests that the increased adoption of remote and hybrid work arrangements reduces the necessity for daily travel to central employment locations, while simultaneously expanding the feasible residential search area (Delventhal et al., 2022; Davis et al., 2024). In this context, households may opt to relocate farther from Toronto to obtain more affordable or larger housing, while still engaging in the metropolitan labor market through reduced commuting frequency. Consequently, the decline in the number of lone-driver commuters should not be interpreted as evidence of weakening economic integration. Instead, it may reflect evolving workplace practices that have decoupled residential distance from daily commuting requirements, thereby allowing housing market adjustments to occur without corresponding increases in automobile travel.

The findings collectively demonstrate that the period from 2016 to 2021 was characterized by significant regional integration of the housing market, accompanied by evolving commuting patterns. The marked increase in dwelling values in both metropolitan and peripheral areas indicates an expansion of housing demand throughout southern Ontario. Simultaneously, the persistent strong spatial gradients in housing values underscore that proximity to Toronto continued to confer a substantial price premium. Furthermore, the widespread increase in the owner shelter cost burden across nearly the entire study region suggests that affordability pressures became more pervasive, extending beyond the metropolitan core. These trends collectively support the hypothesis that rapidly escalating housing prices influence residential location decisions within a broad regional housing system.

The spatial autocorrelation results substantiate the broader conclusion that housing price appreciation, housing affordability, and Toronto-oriented commuting are interrelated yet spatially distinct aspects of metropolitan transformation. In 2016, housing values exhibited geographic clustering, while the shelter-cost burden demonstrated the strongest spatial dependence, indicating that affordability pressures were concentrated among clusters of neighbouring communities rather than being randomly distributed across southern Ontario. By 2021, the attenuation of global clustering in dwelling values and shelter burden suggests that housing-market pressures had become more geographically dispersed. Nonetheless, Local Moran's I results identified persistent areas of similarity and spatial contrast, indicating that this regionalization was uneven. In line with Anselin's (1995) distinction between global and local spatial association, these findings imply an increasingly interconnected housing market with localized affordability and price submarkets.

Toronto-bound car commuting exhibited no significant global spatial autocorrelation in either census year, despite the presence of localized commuter outliers. Additionally, the regression residuals for both the commuting and dwelling-value models indicated minimal remaining spatial dependence. Conversely, the shelter-cost burden demonstrated substantial residual spatial autocorrelation, particularly in 2016, suggesting that distance and dwelling value alone do not fully account for the geography of housing affordability. Collectively, these findings imply that housing price appreciation was linked to an expansion of Toronto's regional housing market. However, this spatial restructuring did not result in a uniform increase in automobile commuting. Instead, housing-market integration appears to have extended over a broader geographic area, while Toronto-bound car commuting remained distinctly influenced by distance and local conditions. The shelter-cost burden was more spatially localized, indicating that affordability is influenced by additional neighborhood and regional factors beyond dwelling prices alone. Therefore, the spatial analysis corroborates the paper's central conclusion that rising housing values contributed to a more geographically extensive metropolitan housing system, while changes in commuting and affordability exhibited more complex and spatially heterogeneous patterns.

This study posits that average dwelling values provide a more insightful measure of metropolitan housing-market dynamics compared to average shelter costs, especially in the context of residential decentralization and commuting patterns. Dwelling values consistently exhibited stronger correlations with distance from Toronto and more effectively depicted the spatial distribution of housing appreciation. In contrast, shelter costs explained significantly less variation, likely because they reflect the financial situations of current homeowners rather than the market conditions encountered by new buyers. This distinction is consistent with theoretical and empirical research suggesting that residential location decisions are primarily influenced by housing acquisition costs rather than the average ongoing expenses of existing owners (So et al., 2001; Van Nieuwerburgh, 2023).

## 6. Managerial and Policy Implications

The findings demonstrate that the rapid appreciation of housing was not limited to Toronto but extended throughout southern Ontario, with significant increases in average dwelling values observed in communities up to approximately 190 km from the metropolitan core. Concurrently, the proportion of owner households allocating more than 30% of their household income to shelter rose across nearly the entire study region. These results imply that housing affordability pressures have become regional rather than solely metropolitan in scope. Therefore, housing policies that focus exclusively on increasing supply within Toronto are unlikely to comprehensively address the affordability challenges. It is recommended that provincial and regional planning authorities implement coordinated housing strategies that expand housing supply across the broader Greater Golden Horseshoe, while enhancing the diversity of housing types available within both established urban centers and rapidly growing peripheral communities.

The study consistently finds that average dwelling value demonstrates a more robust and systematic relationship with both the distance from Toronto and commuting patterns compared to average shelter costs. This distinction holds significant policy implications, as average shelter costs primarily reflect the financial situations of current homeowners, whereas dwelling values indicate the market conditions encountered by prospective buyers. Consequently, housing policies that focus solely on monthly affordability measures may underestimate the challenges faced by first-time buyers and households attempting to enter higher-cost housing markets. By monitoring changes in dwelling values alongside traditional affordability indicators, policymakers could gain a more comprehensive understanding of emerging housing market pressures and enhance the timing of interventions aimed at increasing housing accessibility.

The findings indicate a significant interrelation between housing appreciation and commuting patterns. Despite a substantial increase in dwelling values across peripheral communities, commuter volumes oriented towards Toronto remained strongly correlated with distance, while also showing signs of spatial restructuring between 2016 and 2021. These results underscore the necessity of integrating transportation planning with housing policies. It is essential that residential development in peripheral communities be accompanied by investments in regional transportation infrastructure, such as commuter rail, regional transit, and highway improvements, to ensure continued accessibility to metropolitan employment centers. Coordinated land use and transportation planning would decrease reliance on private automobiles and support sustainable regional growth.

The examination of housing-value gradients and commuter distributions reveals that Toronto's functional labor market extends well beyond the city's administrative boundaries. Communities located more than 100 km from Toronto continue to maintain economic connections with the metropolitan area, despite significant differences in housing prices. These findings imply that employment accessibility should be assessed from a regional perspective rather than being restricted to individual municipal jurisdictions. Therefore, provincial agencies and regional planning organizations should collaborate on infrastructure investment, housing development, and economic strategies across municipalities to more accurately reflect the integrated nature of southern Ontario's housing and labor markets.

The concurrent rise in widespread housing appreciation and varied changes in commuter volumes indicates that residential decentralization is increasingly aligned with a decrease in commuting frequency. This trend aligns with the growth of remote and hybrid work arrangements, as documented in recent urban economic studies (Davis et al., 2022; Davis et al., 2024; Van Nieuwerburgh, 2023). Consequently, transportation demand forecasting and infrastructure planning should incorporate evolving workplace practices rather than solely relying on historical commuting patterns. Assumptions based on five-day office attendance may lead to an overestimation of future peak-period travel demand and an underestimation of the sustained demand for housing in more remote areas.

Numerous peripheral municipalities have witnessed housing appreciation rates that are comparable to or exceed those within the Greater Toronto Area. This outward shift in housing demand indicates that communities previously deemed relatively affordable may face heightened pressure on local housing markets, infrastructure, and public services. Municipal governments should prepare for ongoing population growth by proactively expanding service capacity, transportation networks, schools, healthcare facilities, and other community infrastructure. Concurrently, policies promoting mixed-density residential development may assist in moderating future housing price increases while maintaining affordability for local residents.

The analyses reveal that average dwelling values and the proportion of households allocating more than 30% of their income to shelter expenses represent distinct aspects of housing affordability. Dwelling values indicate obstacles to entering the ownership market, whereas the shelter cost burden assesses the ongoing financial strain on current homeowners. The observed increase in shelter cost burden across nearly all communities suggests that affordability pressures intensified regionally between 2016 and 2021, even in areas where dwelling values remained below those in Toronto. Therefore, housing policy evaluations should integrate both market-based indicators, such as dwelling values, and household affordability measures, such as shelter-cost burden, to offer a more comprehensive assessment of housing market conditions.

The findings collectively suggest that rapid housing appreciation, the rising cost burden of owner shelter, and changing commuting patterns are interconnected aspects of metropolitan restructuring, rather than isolated occurrences. Consequently, effective policy responses necessitate coordinated decision-making across the domains of housing, transportation, land-use planning, and regional economic development. By acknowledging that housing and labor markets are increasingly functioning on a regional scale, policymakers can formulate more integrated strategies that enhance housing affordability while ensuring efficient access to employment throughout the Greater Toronto region.

## 7. Directions for future research

Future research should build upon this study by employing longitudinal household-level data to directly investigate residential mobility, housing-market participation, and commuting decisions. While analyses at the municipality level reveal regional spatial patterns, the use of household microdata would mitigate ecological inference and enable a more robust evaluation of how variations in housing costs and employment conditions affect relocation and commuting. Additionally, incorporating socioeconomic and demographic characteristics—such as income, occupation, household composition, housing tenure, mortgage conditions, and local labor-market characteristics—would further clarify the roles of housing affordability, employment opportunities, and household characteristics in metropolitan decentralization.

Future research should methodologically integrate spatial econometric techniques and more comprehensive accessibility metrics. Evaluating residual spatial dependence could guide the selection of spatial lag, spatial error, or geographically varying models. Additionally, factors such as travel time, congestion, transit availability, network connectivity, and employment accessibility may offer more significant insights into metropolitan accessibility than mere distance. Furthermore, extending the analysis to include public transit, active transportation, ridesharing, and hybrid commuting would yield a more thorough evaluation of the impact of residential decentralization on transportation demand.

Extending both the temporal and geographic scope is of equal significance. Incorporating additional census periods and administrative data may ascertain whether the changes observed between 2016 and 2021—especially those associated with COVID-19 and the rise of remote and hybrid work—constitute temporary disruptions or indicate longer-term structural transformations. Comparative analyses of other Canadian and international metropolitan regions could determine whether Toronto's nonlinear housing and commuting gradients are generalizable and whether their spatial extent varies according to metropolitan size, transportation infrastructure, and housing market conditions.

Future research should transition from descriptive associations to more robust causal identification. Employing longitudinal and quasi-experimental methodologies, such as difference-in-differences, instrumental-variable techniques, and natural experiments related to transportation investments, zoning reforms, or housing-policy modifications, could more effectively isolate the impacts of housing-market appreciation from simultaneous demographic, economic, and workplace changes. These methodological advancements would collectively enhance the understanding of the interactions among housing affordability, residential mobility, accessibility, and commuting, thereby improving the evidence base for integrated metropolitan housing and transportation planning.

**8. Limitations**

The findings must be considered with regard to several methodological and data constraints. First, the analysis conducted at the municipality and census metropolitan area levels identifies aggregate spatial relationships rather than household-level behaviors, thus posing a risk of ecological fallacy (Robinson, 1950). Longitudinal data at the household level could more effectively examine whether variations in housing costs affect residential mobility and commuting decisions (So et al., 2001). Second, since regressions were estimated separately for 2016 and 2021, the analysis highlights differences in spatial relationships rather than longitudinal effects. Panel methods could more accurately isolate temporal changes while accounting for time-invariant municipal characteristics (Baltagi, 2021). The relatively small sample size also limits statistical power and heightens sensitivity to influential observations (Kutner et al., 2005), particularly in Toronto, which significantly differs from other study areas in terms of dwelling values and commuter volumes. Therefore, employing influence diagnostics and conducting sensitivity analyses that exclude Toronto would enhance model validation (Cook, 1977; Belsley et al., 1980). Third, the models fail to incorporate factors that may affect residential location and commuting, such as income, employment opportunities, housing supply, mortgage rates, transportation accessibility, amenities, and planning policies. As a result, the estimated distance effects may partially reflect unobserved characteristics and should be interpreted as associations rather than independent causal effects. Multivariable or structural models could more accurately account for these relationships (Kline, 2023). Fourth, although quadratic regression captured nonlinear spatial patterns, more flexible approaches—such as generalized additive models, restricted cubic splines, and LOESS—could assess whether the observed relationships depend on the assumed functional form (Hastie & Tibshirani, 1990; Harrell, 2015). Fifth, commuter counts were highly skewed and influenced by differences in population size; modeling commuter rates or employing generalized linear count models with appropriate exposure measures could offer valuable robustness tests (Cameron & Trivedi, 2013). Finally, the period from 2016 to 2021 includes the COVID-19 pandemic and the rapid expansion of remote and hybrid work, complicating the separation of housing-market effects from concurrent changes in employment and commuting behavior (Delventhal et al., 2022; Davis et al., 2024). Future research utilizing longitudinal or quasi-experimental designs could more effectively differentiate housing-price effects from simultaneous economic, transportation, and behavioral changes, thereby strengthening causal inference (Angrist & Pischke, 2009).

## 9. Funding

This research did not receive any specific grant from funding agencies in the public, commercial, or not-for-profit sectors.

## 10. Conflict of interest

The authors declare no conflicts of interest.

.